\documentclass[12pt]{article}

\usepackage{epsf,amsfonts,hyperref}
\usepackage[dvips]{color}
\usepackage{graphicx}
\renewcommand{\appendix}[1]{
    \addtocounter{section}{1}
    \setcounter{equation}{0}
    \renewcommand{\thesection}{\Alph{section}}
    \section*{Appendix \thesection\protect\indent #1}
    \addcontentsline{toc}{section}{Appendix \thesection\ \ \ #1}
}
\newcommand\encadremath[1]{\vbox{\hrule\hbox{\vrule\kern8pt
\vbox{\kern8pt \hbox{$\displaystyle #1$}\kern8pt}
\kern8pt\vrule}\hrule}}
\def\enca#1{\vbox{\hrule\hbox{
\vrule\kern8pt\vbox{\kern8pt \hbox{$\displaystyle #1$}
\kern8pt} \kern8pt\vrule}\hrule}}

\newcommand\figureframex[3]{
\begin{figure}[bth]
\hrule\hbox{\vrule\kern8pt
\vbox{\kern8pt \vbox{
\begin{center}
{\mbox{\epsfxsize=#1.truecm\epsfbox{#2}}}
\end{center}
\caption{#3}
}\kern8pt}
\kern8pt\vrule}\hrule
\end{figure}
}
\newcommand\figureframey[3]{
\begin{figure}[bth]
\hrule\hbox{\vrule\kern8pt
\vbox{\kern8pt \vbox{
\begin{center}
{\mbox{\epsfysize=#1.truecm\epsfbox{#2}}}
\end{center}
\caption{#3}
}\kern8pt}
\kern8pt\vrule}\hrule
\end{figure}
}

\renewcommand{\thesection}{\arabic{section}}

\makeatletter
\@addtoreset{equation}{section}
\makeatother
\newtheorem{theorem}{Theorem}[section]

\newtheorem{remark}{Remark}[section]
\newtheorem{proposition}{Proposition}[section]
\newtheorem{lemma}{Lemma}[section]
\newtheorem{corollary}{Corollary}[section]
\newtheorem{definition}{Definition}[section]
\def\br{\begin{remark}\rm\small}
\def\er{\end{remark}}
\def\bt{\begin{theorem}}
\def\et{\end{theorem}}
\def\bd{\begin{definition}}
\def\ed{\end{definition}}
\def\bp{\begin{proposition}}
\def\ep{\end{proposition}}
\def\bl{\begin{lemma}}
\def\el{\end{lemma}}
\def\bc{\begin{corollary}}
\def\ec{\end{corollary}}
\def\beaq{\begin{eqnarray}}
\def\eeaq{\end{eqnarray}}

\newcommand{\eq}[1]{Eq.~(\ref{#1})}

\newcommand{\beq}{\begin{equation}}
\newcommand{\eeq}{\end{equation}}
\newcommand{\bea}{\begin{eqnarray}}
\newcommand{\eea}{\end{eqnarray}}

\renewcommand{\and}{{\qquad {\rm and} \qquad}}

\newcommand{\virg}{{\qquad , \qquad}}

 \newcommand{\Tr}{{\,\rm Tr}\:}

\newcommand{\Res}{\mathop{\,\rm Res\,}}

\newcommand{\td}[1]{{\tilde{#1}}}

\renewcommand{\l}{\lambda}
\newcommand{\om}{\omega}

\newcommand{\ee}[1]{{{\rm e}^{#1}}}

\newcommand{\Pint}{{\int\kern -1.em -\kern-.25em}}

\renewcommand{\Re}{{\mathrm{Re}}}
\renewcommand{\Im}{{\mathrm{Im}}}

\renewcommand{\l}{\lambda}

\newcommand{\ovl}{\overline}

\newcommand{\acycle}{{\cal A}}
\newcommand{\bcycle}{{\cal B}}

\newcommand{\curve}{{\cal E}}

\newcommand{\Jac}{{\mathbb J}}

\newcommand{\spcurve}{{\cal S}}
\renewcommand{\curve}{{\cal C}}

\newcommand{\genus}{{\mathfrak{g}}}
\newcommand{\CYX}{\mathfrak X}
\newcommand{\NN}{{\cal N}}

\begin{document}
\sloppy


\pagestyle{empty}
\hfill IPHT-T09/196

\hfill CERN-PH-TH-2009-230

\addtolength{\baselineskip}{0.20\baselineskip}
\begin{center}
\vspace{26pt}
{\large \bf {Geometrical interpretation of the topological recursion, and integrable string theories. }}
\newline
\vspace{26pt}

{\sl Bertrand  Eynard}\hspace*{0.05cm}\footnote{ E-mail: bertrand.eynard@cea.fr },
{\sl Nicolas  Orantin}\hspace*{0.05cm}\footnote{ E-mail: nicolas.orantin@cern.ch }\\
\vspace{6pt}
Institut de Physique Th\'eorique,\\
CEA, IPhT, F-91191 Gif-sur-Yvette, France,\\
CNRS, URA 2306, F-91191 Gif-sur-Yvette, France.\\
\vspace{6pt}
Theory division, CERN\\
CH-1211 Geneva 23, Switzerland.\\

\end{center}

\vspace{20pt}
\begin{center}
{\bf Abstract:}
Symplectic invariants introduced in \cite{EOFg} can be computed for an arbitrary spectral curve. For some examples of spectral curves, those invariants can solve loop equations of matrix integrals, and many problems of enumerative geometry like maps, partitions, Hurwitz numbers, intersection numbers, Gromov-Witten invariants... The problem is thus to understand what they count, or in other words, given a spectral curve, construct an enumerative geometry problem.
This is what we do in a  semi-heuristic approach in this article.
Starting from a spectral curve, i.e. an integrable system, we use its flat connection and flat coordinates, to define a family of worldsheets, whose enumeration is indeed solved by the topological recursion and symplectic invariants.
In other words, for any spectral curve, we construct a corresponding string theory, whose target
space is a submanifold of the Jacobian.
\end{center}

\vspace{26pt}
\pagestyle{plain}
\setcounter{page}{1}

\tableofcontents

\section{Introduction}

Topological String Theories aim at addressing the question of "counting" how many Riemann surfaces (worldsheets) with given boundary conditions, can be embedded into a given target space.
Witten suggested \cite{Witten} an underlying string field theory, in which worldsheets are obtained by gluing some basic building blocks. For example in Teichm\"uller theory, building blocks are "pairs of pants".
In Kontsevich's approach, building blocks are cylinders glued along a ribbon graph, and this idea has then given many variants.

Recently, it was suggested by BKMP \cite{BKMP}, that Gromov-Witten amplitudes of the type A topological strings in a toric CY 3-fold target space $\CYX$, coincide with the "symplectic invariants" (introduced in \cite{EOFg}) of the spectral curve $\spcurve_{\td \CYX}$ of the mirror $\td{\CYX}$ of the target space $\CYX$:
\beq
\encadremath{
\hbox{BKMP conjecture:}\qquad \quad {\rm GW}_g(\CYX) \stackrel{?}{=}\,\, F_g(\spcurve_{\td{\CYX}}).
}\eeq
The main interest of that conjecture, is that the right hand side, i.e. the symplectic invariants, is  much easier to compute than the left hand side for given genus.

\medskip
Here, we shall study this claim, and try to understand its geometric meaning.

\medskip
In fact, we shall work backwards, and starting from a spectral curve $\spcurve$, we shall try to construct a "string theory" whose partition function is given by the symplectic invariants $F_g$'s.

\smallskip
The basic idea, is that out of a spectral curve, we can construct an integrable system, and in particular flat connections, and a system of action-angle variables. For every initial condition, the angle variables of an integrable system, follow a uniform linear motion in time, which means that they generate a 1-dimensional manifold in the phase space, and, because of conformal invariance, time is a complex variable, it means a 1-dimensional complex manifold, i.e. a Riemann surface embedded in the phase space (the phase space of action angle variables has a toric symmetry).
Moreover, since the motion is uniform, the angle coordinate is a flat coordinate almost everywhere on each such Riemann surface, and this gives a natural foliation on all surfaces. These surfaces can thus be cut into "propagators" and cylinders in a unique way.
This immediately implies that the generating functions which count such Riemann surfaces of a given topology, do satisfy the topological recursions of \cite{EOFg}, and thus they are the symplectic invariants.

\medskip
{\bf Outline}

In section 2, we recall the definition of the symplectic invariants as well as some of their properties. We also recall that they have a diagrammatic representation, which resembles very strongly what could be expected for a string field theory.
\smallskip

In section 3, we consider an arbitrary integrable system, we recall the notion of "action-angle" flat coordinates, and how action angle coordinates can be used to generate worldsheets of a string theory.
In other words, we define an adhoc string theory attached to an integrable system.

\smallskip

In section 4, we define moduli spaces of worldsheets of given topologies and brane boundary conditions, and we show that worldsheets can be decomposed into propagators and cylinders.
This induces a decomposition of moduli spaces of worldsheets into cells labeled by graphs.

\smallskip

In section 5, we translate the decomposition of moduli spaces in terms of string amplitudes, in local patch coordinates.
A consequence is that, after Laplace transforms, string amplitudes obey the topological recursion of \cite{EOFg}. In particular, closed string amplitudes of genus $g$ are the symplectic invariants $F_g$.

\smallskip

In section 6, we rewrite amplitudes in terms of intrinsic geometry of the spectral curve. That allows to identify the spectral curve with the disc amplitude, and the Bergman kernel with the cylinder amplitude.

\smallskip

In section 7, we show how expanding the generating functions in terms of other formal variables, through Lagrange inversion formula, can give to many combinatorial identities. This generalizes Cut and Join equations and ELSV formulae. Indeed, expansions near branchpoints of the spectral curve can always be written in terms of intersection numbers, and expansions near singularities of the spectral curve can be written in terms of winding numbers.

\smallskip

In section 8 we discuss the case of toric CY target spaces leading to a geometrical interpretation of BKMP conjecture.

\smallskip
Section 9 is the conclusion.

\section{Symplectic invariants of a spectral curve}

Let $\spcurve$ be a spectral curve, i.e. it is the data of a compact Riemann surface $\curve$ of genus\footnote{the genus $\genus$ of the spectral curve has nothing to do with the genus $g$ of worldsheets studied further.} $\genus$, and two analytical functions $x$ and $y$ on $\curve$, or on some open domain of $\curve$:
\beq\label{eqdefspcurve}
\spcurve=(\curve,x,y).
\eeq

Typically, in topological strings, $\curve$ is an algebraic curve of equation $H(x_+,x_-)=0$ where $H$ is some polynomial, and $x=\ln{(x_+)}, y=\ln{(x_-)}$.

Notice that $(\curve,x)$ is a Hurwitz space, i.e. the data of a compact Riemann surface $\curve$ together with a projection $x:\curve\to \mathbb CP^1$, which realizes $\curve$ as a branched covering of $\mathbb CP^1$, with a
branching structure given by the zeroes and poles of the differential $dx$:  the branchpoints are the zeroes and poles of $dx$, and the monodromies are the orders of the zeroes and poles of $dx$.

From now on, we assume that $dx$ is meromorphic, and all zeroes of $dx$ are simple (this is the case for spectral curves of topological strings, and also for spectral curves of matrix models counting discrete surfaces).

\bigskip
{\bf $\bullet$ Bergman kernel}

Then we define a Bergman kernel $B(z_1,z_2)$ on ${\cal C}$, i.e. a 2nd kind meromorphic symmetric 2-form on ${\cal C}$ having a double pole with vanishing residue at $z_1=z_2$ and no other pole. It is normalized by requiring that near $z_1=z_2$, in any parametrization $\xi(z)$ it behaves like:
\beq\label{eqdefBergman}
B(z_1,z_2)=B(z_2,z_1) \mathop{{\sim}}_{z_1\to z_2}\, {d\xi(z_1)\,d\xi(z_2)\over (\xi(z_1)-\xi(z_2))^2}\,\, + {\rm regular} .
\eeq
As defined here, the Bergman kernel is not unique, one may add to it any combination of holomorphic forms, i.e.
differential forms without poles.

For a given symplectic basis of noncontractible cycles on ${\cal C}$, i.e. $2\genus$ cycles $\left\{\acycle_i,\bcycle_i\right\}_{i=1}^\genus$, satisfying:
$$
\acycle_i \cap \bcycle_j = \delta_{i,j},
$$
one can define a basis of holomorphic forms $\{du_i\}_{i=1}^\genus$ normalized by
\beq\label{eqdefholomorphic}
\oint_{\acycle_i} du_j(z) = \delta_{i,j} \qquad \hbox{and} \qquad \oint_{\acycle_i} du_j(z) = \tau_{i,j}
\eeq
where $\tau$ is the Riemann matrix of periods (see for instance \cite{Farkas,Fay}).
One can then parameterize the holomorphic deformations of the Bergman kernel with a
symmetric matrix $\kappa$ of size $\genus\times\genus$, and one may consider the Bergman kernel shifted by a combination of holomorphic forms, as a new admissible Bergman kernel:
\beq\label{eqBpluskappa}
B(z_1,z_2)\to B(z_1,z_2) + 2i\pi\,\, \sum_{i,j=1}^\genus\,\, du_i(z_1)\,\kappa_{i,j}\,du_j(z_2) .
\eeq
A choice of $\kappa$ is more or less equivalent to a choice of a symplectic basis of cycles
$\acycle_i(\kappa),\bcycle_i(\kappa), i=1,\dots,\genus$ on which the Bergman kernel is normalized by
$$
\oint_{z_1 \in \acycle_i(\kappa)} B(z_1,z_2) = 0.
$$
From now on, let us assume that we have chosen a Bergman kernel, or in other words a matrix $\kappa$, or in other words a basis of cycles.


\bigskip
{\bf $\bullet$ Branchpoints and conjugated points}

The branchpoints $a_i$ are the points with a "vertical tangent", i.e. the zeroes of $dx$:
\beq
dx(a_i)=0.
\eeq
We assume that all branchpoints are regular, i.e. they are simple zeroes of $dx$, and they are not zeroes of $dy$
nor poles of $y$.
This means that near a branchpoint $z_i$, the curve behaves like a square-root:
\beq
y(z) \mathop{\sim}_{z\to a_i}\,\, y(a_i) + C_i \sqrt{x(z)-x(a_i)} + \dots
\eeq
This also means that in a small vicinity of $a_i$, there is a unique point $\bar{z}\neq z$ in the same vicinity of $a_i$ such that:
\beq
x(\bar{z})=x(z).
\eeq
$\bar{z}$ is called the conjugated point of $z$.
It is defined only locally near branchpoints, and it is not necessarily defined globally\footnote{A notable exception is the case of hyperelliptical surfaces, where $z\to\bar{z}$ is the hyperelliptical involution and is defined globally.}.

\figureframex{8}{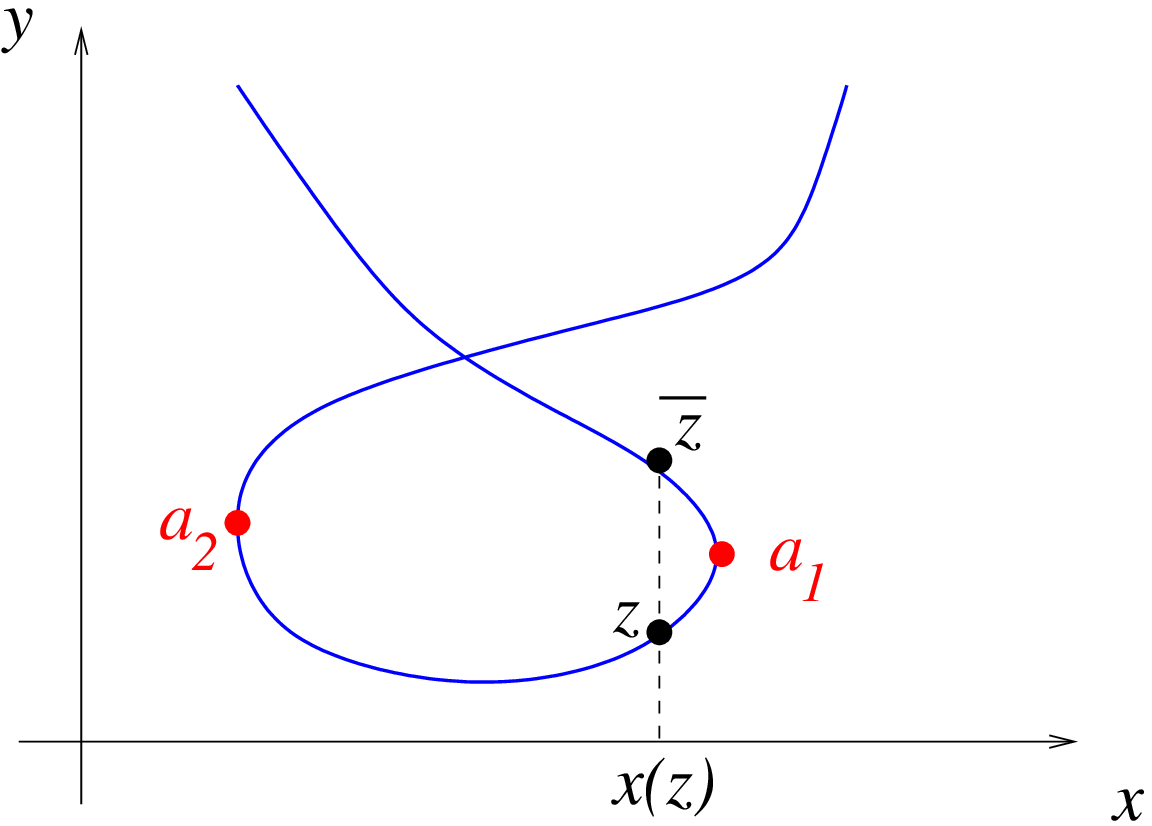}{Branchpoints $a_i$ are points with a vertical tangent. Near a branch point $a_i$, there are two branches coming together. If $z$ is on one branch, we call $\bar z$, the point on the other branch with the same $x$ projection $x(\bar z)=x(z)$. Notice that $\bar z$ is not globally defined, it is defined only locally near branchpoints. For example, if $z$ moves from $a_1$ to $a_2$, the analytic continuation of $\bar z$ would take the wrong branch.}

\bigskip
{\bf $\bullet$ Recursion kernel}

We define the recursion kernel:
\beq\label{eqdefK}
K(z_1,z) = K(z_1,\bar{z}) = {-\,\int_{\bar{z}}^z\,\, B(z_1,z')\over 2(y(z)-y(\bar{z}))\, dx(z)}.
\eeq
It is a 1-form in $z_1$, defined globally on $z_1\in{\cal C}$.
It is the inverse of a 1-form in $z$, defined only locally near branchpoints.

It has the property that it has a simple pole at branch points, and near $z\to a_i$ it behaves like:
\beq
K(z_1,z)  \mathop{\sim}_{z\to a_i}\,\, -\,{B(z_1,z)\over 2\,dx(z)\,dy(z)} + {\rm regular}.
\eeq

\bigskip
{\bf $\bullet$ Symplectic invariants}

Then, following \cite{EOFg}, we define a sequence of symmetric meromorphic $n$-forms, called $\om_n^{(g)}$ for every $n$ and $g$ integers, by the following recursion (often called "topological recursion"):
\beq
\om_1^{(0)}(z) = - y(z)\,dx(z)
\eeq
\beq
\om_2^{(0)}(z_1,z_2) = B(z_1,z_2)
\eeq
\bea\label{toprecdef}
\om_{n+1}^{(g)}(z_1,\dots,z_n,z_{n+1})
&=& \sum_i \Res_{z\to a_i}\, K(z_{1},z)\, \Big[ \om_{n+2}^{(g-1)}(z,\bar{z},z_2,\dots,z_{n+1}) \cr
&& + \sum_{h=0}^g\,\sum'_{I\subset J}\, \om_{1+|I|}^{(h)}(z,I)\,\om_{1+n-|I|}^{(g-h)}(\bar{z},J\backslash I) \Big] \cr
\eea
where $J$ is a collective notation $J=\{z_2,\dots,z_{n+1}\}$, and $\sum'$ means that we exclude the terms $(h,I)=(0,\emptyset),(g,J)$.

We also define if $g\geq 2$:
\beq
F_g \equiv \om_0^{(g)} = {1\over 2-2g}\, \sum_i \Res_{z\to a_i}\, \om_1^{(g)}(z)\, \Phi(z)
\eeq
where $\Phi$ is any analytical function defined in the vicinity of branchpoints such that $d\Phi=ydx$.
There are also definitions for $F_0$ and $F_1$, and we refer the reader to \cite{EOFg} for those two cases. $F_0$ is often called the prepotential.

\smallskip
All the $\om_n^{(g)}$'s with $2-2g-n<0$ are called stable, and the others are called unstable. The only unstable ones are $F_0, \om_1^{(0)}, \om_2^{(0)}, F_1$.

\smallskip

Although the definition doesn't look symmetric, every $\om_n^{(g)}$ is a symmetric $n$-form.
Stable $\om_n^{(g)}$'s have poles only at branchpoints, of order at most $6g-4+2n$, and with vanishing residues.

\subsection{Some properties of symplectic invariants}

\bigskip
{\bf $\bullet$ Rescaling}

Under a rescaling $y\to\l y$, we have (if $2-2g-n\neq 0$):
\beq
\om_n^{(g)} \to \l^{2-2g-n}\, \om_n^{(g)},
\eeq
and in particular (for $g\neq 1$):
\beq
F_g \to \l^{2-2g}\, F_g.
\eeq
In particular, this implies that $F_g$ is invariant under the parity transformation $y\to -y$.

\bigskip
{\bf $\bullet$ Symplectic invariance}

If two spectral curves $\spcurve=({\cal C},x,y)$ and $\td\spcurve=(\td{\cal C},\td{x},\td{y})$ are such that there is an analytical bijection from ${\cal C}\to\td{\cal C}$ which conserves the symplectic form in $\mathbb C\times \mathbb C$:
\beq
dx\wedge dy = d\td{x}\wedge d\td{y},
\eeq
then we have (for $g\geq 2$):
\beq
F_g(\spcurve) = F_g(\td\spcurve).
\eeq
This is why we call the $F_g(\spcurve)$ the symplectic invariant of degree $2-2g$.

\bigskip
{\bf $\bullet$ Other properties}

There are many other properties, for instance concerning modularity, infinitesimal variations of spectral curve, singular limits, and integrability.
For example, for any spectral curve, the $F_g$'s satisfy holomorphic anomaly equations \cite{eynhaeq}.

Also, it turns out that the $F_g$'s allow to construct a Tau-function, and an integrable system associated to $\spcurve$.

All those properties can be found in \cite{EOFg,EOFgreview}.

\subsection{String Field theory: diagrammatical rules}

Let us represent pictorially every $\om_n^{(g)}$ as a surface of genus $g$ with $n$ boundaries labeled by $z_1,\dots, z_n$.
$${\mbox{\epsfxsize=6.2truecm\epsfbox{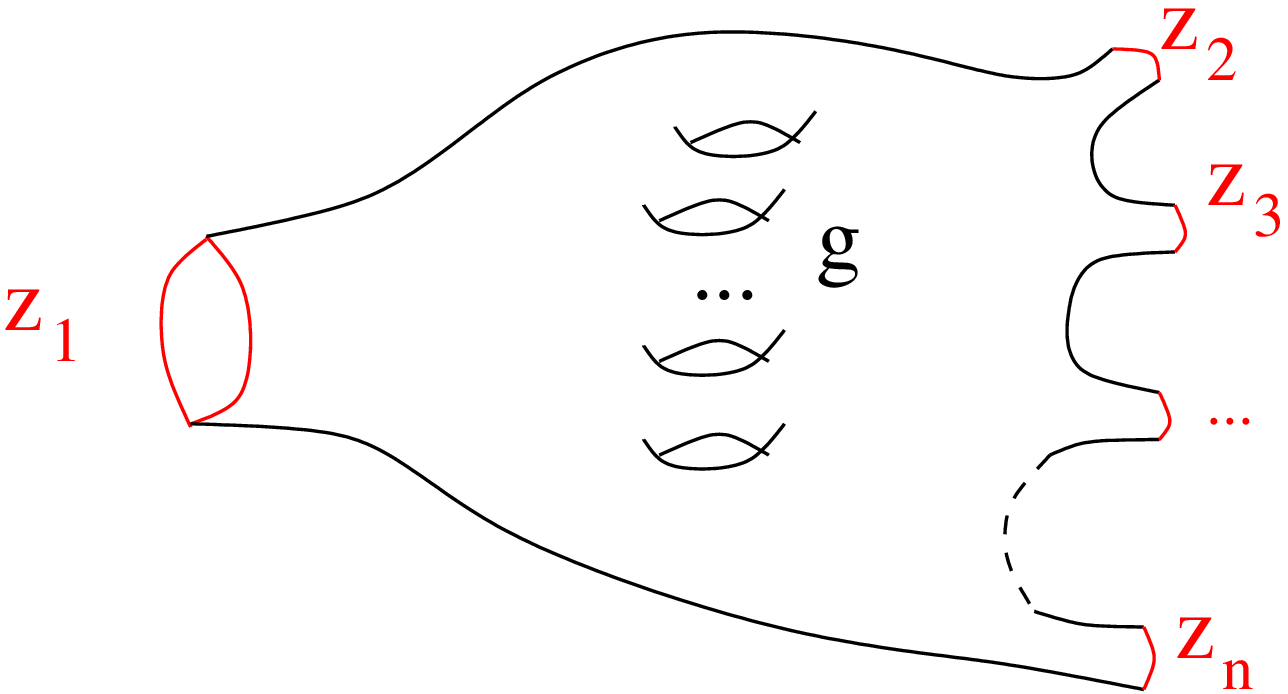}}}$$

Let us represent our two kernels as elementary pieces used to build such surfaces:

\bigskip
{\bf $\bullet$ The "propagator"}
\beq
K(z_1,z) = {\mbox{\epsfxsize=3.2truecm\epsfbox{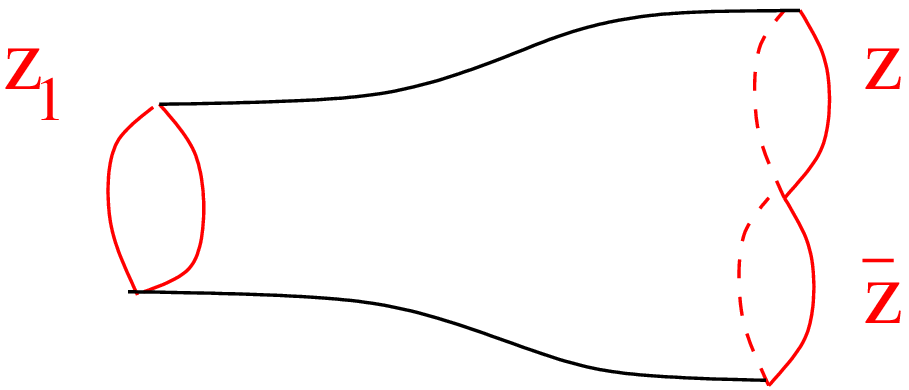}}}
\eeq

\bigskip
{\bf $\bullet$ The "cylinder":} the two point function $\om_2^{(0)}(z_1,z_2)$
\beq
B(z_1,z_2) = {\mbox{\epsfxsize=3.8truecm\epsfbox{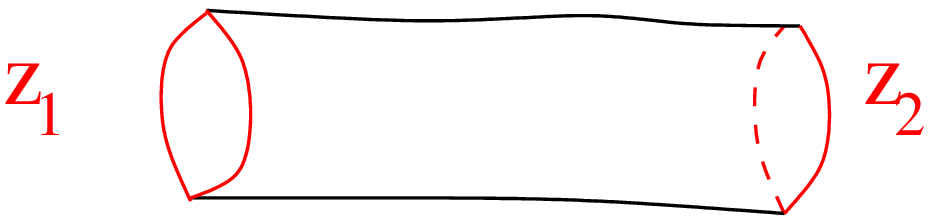}}}
\eeq

\bigskip
{\bf $\bullet$  Recursion formula}: the recursion formula \eq{toprecdef} can be represented as
\beq\label{eqrectopstrft}
{\mbox{\epsfxsize=16.truecm\epsfbox{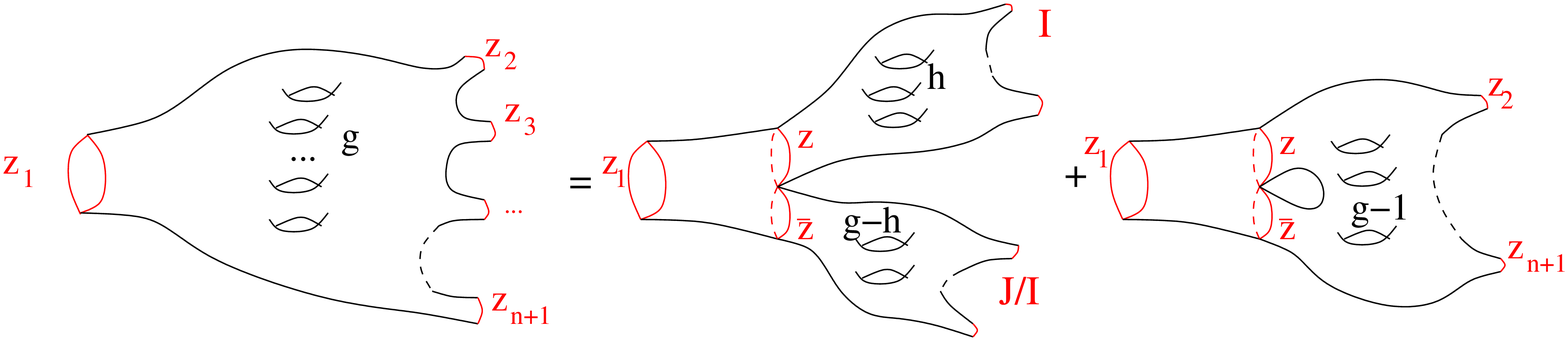}}}
\eeq
where one integrates over the intermediate variable $z$.
This recursion is said to be topological since the surfaces generated in the right hand
side have strictly higher Euler characteristic than the one of the left hand side, and thus this recursion terminates after a finite number of steps equal to minus the Euler characteristics.

In other words, the topological recursion tells us that every surface enumerated by the $\om_n^{(g)}$'s, can be decomposed into $2g-2+n$ propagators and $n+g-1$ cylinders, more or less in a unique way.

\bigskip
{\bf Example:}

For $\om_2^{(1)}$, our recursion formula gives in two iterations:
\bea
{\mbox{\epsfxsize=4.5truecm\epsfbox{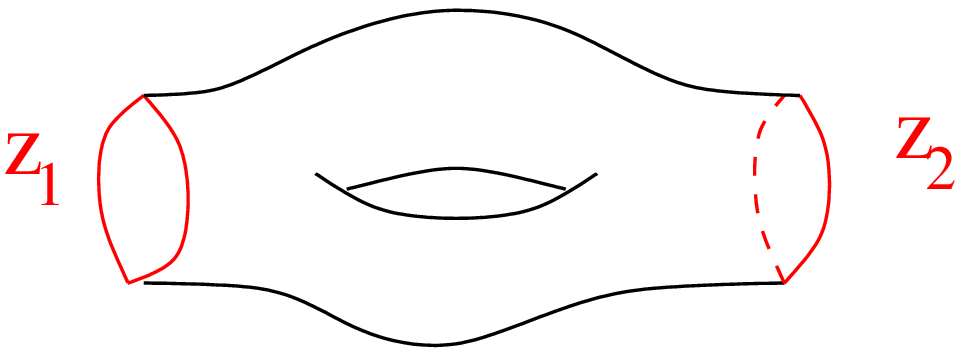}}}
&=& 2{\mbox{\epsfxsize=8.5truecm\epsfbox{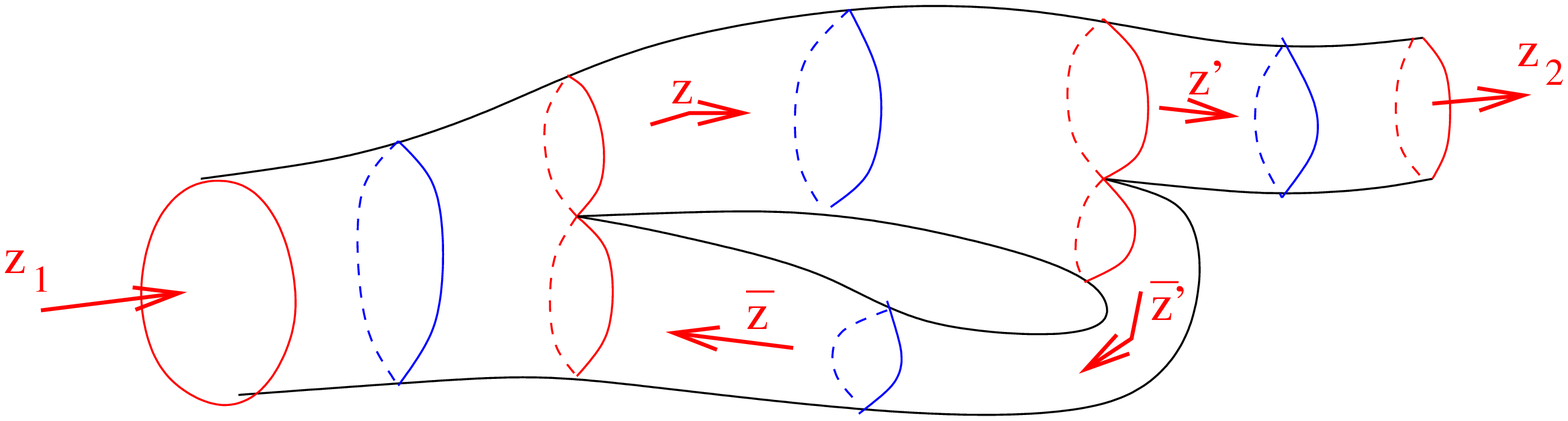}}} \cr
&& +2{\mbox{\epsfxsize=8.5truecm\epsfbox{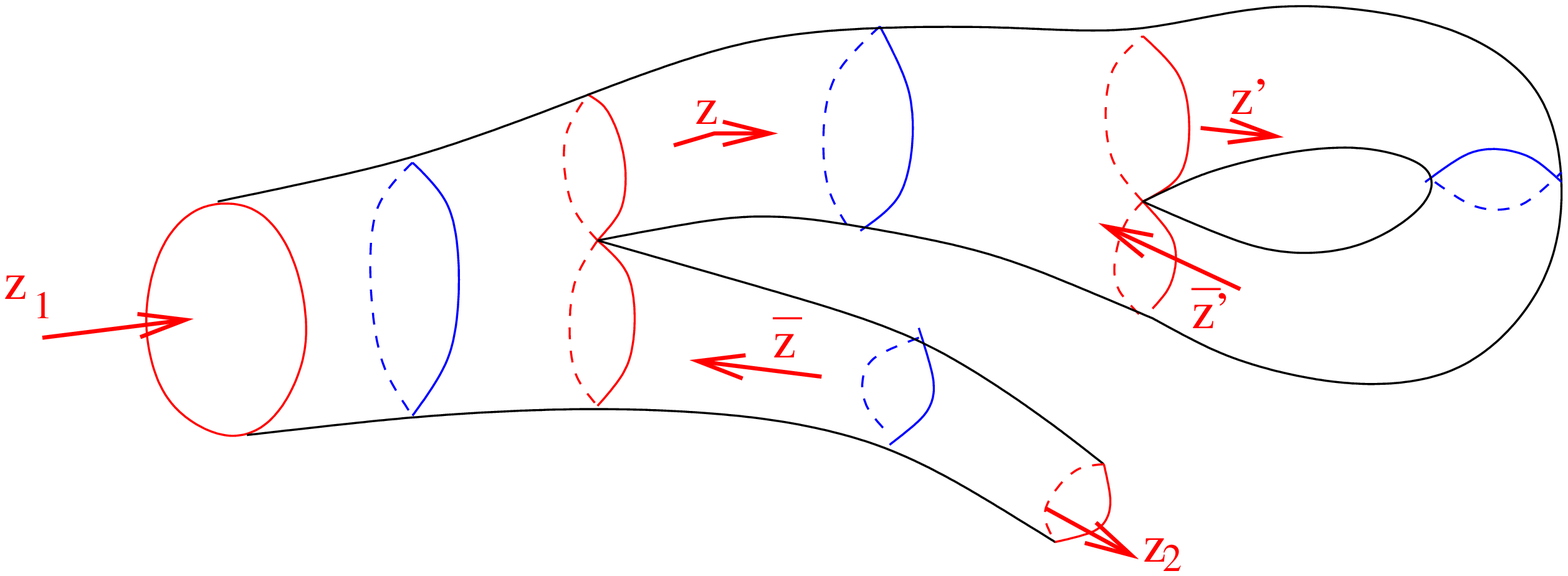}}}\cr
\eea
which is merely the pictorial representation of the following residue formula:
$$
\om_2^{(1)}(z_1,z_2) = 2 \sum_{i,j} \Res_{z \to a_i} \Res_{z' \to a_j} K(z_1,z) K(z,z') \Big[B(z',z_2) B(\bar{z},\bar{z}') + B(\bar{z},z_2) B(z',\bar{z}')\Big].
$$

\bigskip
That diagrammatic representation of the topological recursion, resembles strongly what one could expect to be "string field theory" diagrammatic rules, i.e. how to compute string theory amplitudes by gluing surfaces.

\medskip

In fact it was conjectured by BKMP \cite{BKMP} that Gromov-Witten amplitudes, i.e. topological string theory amplitudes, which enumerate surfaces embedded in a certain target space $\CYX$, should be equal to the symplectic invariants of the spectral curve which is the singular locus in the mirror  of the target space $\CYX$.
Here we are going to try to explain heuristically why the topological recursion indeed counts some string theory amplitude by giving a meaning to the graphical representation of the present section.

But we are going to proceed backwards, i.e. given a spectral curve and its symplectic invariants, we are going to construct a corresponding effective string theory.

\subsection{Flavor of string theory}\label{secintroinstanton}

Just in order to give some intuition of what we do in the next section, we recall in a very sketchy way, the basic ideas underlying string theory.

\medskip
Consider a "space-time" called {\bf target space} $\CYX$, which is a complex manifold of fixed dimension $D$.

A "closed string" is an embedding in $\CYX$ of a circle: $\left\{\vec X(\sigma)\, , \,\,\, \sigma\in [0,2\pi]\right\} \subset \CYX$ with $\vec X(0)=\vec X(2\pi)$, and an open string is an embedding of a segment, i.e. we don't assume $\vec X(0)=\vec X(2\pi)$. From now on, let us consider only closed strings.
As the string  moves with time $T$ in the target space, the history of the string sweeps a surface, called {\bf world sheet} in the target space:
\beq
\vec X(\sigma,T)\in \CYX
\eeq
where $\sigma\in[0,2\pi]$ is the coordinate along the string, and $T\in \mathbb R_+$ is the time coordinate, and $\vec X=(X_1,\dots,X_D)$ is a point of $\CYX$, in a local coordinate system.

String theory amplitudes are obtained by "counting" how many histories can relate an initial state to a final state, i.e. enumerating surfaces having given boundaries.
Surfaces should also be counted with a weight, typically the exponential of some action (for instance Nambu-Gotto's action depends only on the total curvature and area of the surface spanned in the target space).
In other words, we have to perform a functional integral over all coordinates $X_i(\sigma,T)$\footnote{Following our
description, these amplitudes are scattering of closed strings. However, they are referred to as open amplitudes
in the topological string literature since they enumerate surfaces with boundaries, i.e. open surfaces. This name
can also be seen as originating from the open-closed duality obtained by considering $\sigma$ as the time
instead of $T$. The worldsheet is then spanned by open strings ending on some manifolds called Branes.}:
\beq
W({\rm boundaries}) \, ":="\, \int\,\, \prod_{i=1}^D\,{\cal D}[X_i(\sigma,T)] \,\,\, \ee{-{\rm Action}[\vec X(\sigma,T)]}.
\eeq
The "functional measures" ${\cal D}[X_i(\sigma,T)]$ can be quite complicated depending on the geometry of $\CYX$. When $\CYX$ is a submanifold of $\mathbb C^{D+D'}$, the functional integrals can be implemented in terms of standard functional measures in  $\mathbb C^{D+D'}$,  by Lagrange multipliers which enforce $D'$ relationships between the components $X_i, \, i=1,\dots,D+D'$. This procedure leads to a so-called $\sigma$-model
description of this string theory. This is just one representation of that theory among others but it has the
advantage of being sufficiently well known, for some classes of target spaces, to be mapped to some integrable system (see \cite{Dorey} for example).
The boundary conditions can also be implemented by Lagrange multipliers.

\medskip

Now we shall assume that the theory has a conformal invariance property, i.e. that the functional measures and the action, are invariant under conformal reparametrizations of the worldsheet.
In other words, we want to count only once worldsheets which are conformally equivalent.
Choosing one representant per equivalence class is often realized by "gauge fixing", i.e. by introducing "ghosts", but we shall not need it in this article.

We shall only say that $\vec X(\sigma,T)=\vec X(\sigma+iT)$ is an analytical function of a complex variable $t=\sigma+iT$, and conformal reparametrizations, are obtained by changing $t$ to an analytical function of $t$.
In other words, the worldsheet is a 1 dimensional-complex motion $\vec X(t)$ in $\CYX$ with a complex time $t$. Gauge fixing means choosing a complex coordinate  $t$ on each worldsheet.
One of the main points in our article, is to find a "canonical" choice of time coordinate.

\medskip

Then, we shall make another assumption, which is that our action is  integrable.
There are many definitions of integrability. At the classical level, one of them can be phrased like: any motion which extremizes the action
(and then called "classical motion"), has as many conserved quantities as half the dimension of the phase space, and which thus implement the same number of commuting hamiltonian flows, and thus imply a toric symmetry.
Another one, more convenient for us, is that, for every classical motion, there exists a suitable change of variables,
mapping the coordinates $X_i(t), \dot X_i(t)$ to the so-called "action  and angle" variables such that after the change of variables,
the motion is linear at constant speed in a multi-dimensional torus (see \cite{bookintegrable} for an introduction to
integrable systems).
This also provides a natural torus action.

A consequence of integrability and a torus action, is that there is a localization formula, and the functional integral above can be reduced to a sum over only worldsheets which extremize the action. Such extremal worldsheets are often called classical trajectories, classical motions, or instantons.

We thus have:
\beq
W({\rm boundaries}) \, "="\, \sum_{{\rm instantons}} \,\,\, \ee{-{\rm Action}[\vec X(\sigma,T)]}.
\eeq
It just remains to count how many instantons there are, with given boundary conditions.

\medskip

Typically, we shall require not only that boundaries are fixed, but also that the topology of worldsheets be fixed.
Concerning boundary conditions, it is certainly possible to imagine a very large set of possible boundary conditions (modulo conformal reparametrizations again), but we shall consider specific boundary conditions which
can be classified by some moduli and quantum numbers.
In other words, we shall consider only certain types of boundary conditions, often called branes, which can be parametrized by a finite number of complex variables referred to as open moduli.

Finally, that defines a function:
\beq
W_k^{(g)}(z_1,\dots,z_k)
\eeq
which is the amplitude counting worldsheets of genus $g$, with $k$ boundaries parametrized by $k$ open moduli $z_1,\dots,z_k$.

\medskip

This paragraph was only a very sketchy and imprecise introduction to string theory.

Our goal in this article, is to try to understand why those string theory amplitudes are the same as those computed by the symplectic invariants and topological recursion of \cite{EOFg}.

\section{Integrability}

Here, we don't assume to have any string theory, instead we are going to construct one.
Our starting point is a spectral curve as defined by \eq{eqdefspcurve}.

\medskip
Given a spectral curve, it is always possible to construct a classical integrable system (see the reconstruction formula \cite{bookintegrable}). There is not a unique integrable system corresponding to a given spectral curve, but they should all have the same Tau-function.
Let us choose one of them (this arbitrariness should be linked to the choice of framing and background in topological strings, see section \ref{secambiguity}, and in general it is linked to the choice of one hamiltonian  among the family of commuting hamiltonians).

\smallskip

In other words, our starting point is an integrable system, and our goal is to enumerate the classical trajectories given by the equations of motion of this integrable system.


\subsection{Action-angle variables and flat coordinates}

Consider a classical integrable system, with a rather arbitrary target space $\mathfrak X$, with coordinates (in a local patch) $\vec X=(X_1,\dots,X_D)$.
Suppose that the motion $\vec X(t)$ is a solution of the Hamilton's equations of motion of our integrable theory, or in other words it is an extremum of the action (Hamilton-Jacobi equations).

\smallskip

Then, all classical integrable systems have the property that there is (almost everywhere), a canonical change of variables $(\vec X,\partial_t{\vec X})\to (\vec \epsilon, \vec u)$ (called action-angle variables)
which brings the complicated motion $\vec X(t)$ into a linear motion at constant velocity $\vec v$ in the Jacobian (see fig. \ref{figactionangle}):
\beq\label{eqmotionactangle}
\vec u(t)=\vec u(0)+ t\,\vec v
\virg
\vec\epsilon = {\rm constant}.
\eeq
The Jacobian $\Jac$ is a $\genus$-dimensional torus, with some quasi-periodicity properties:
\beq
\vec u \equiv \vec u \,\, {\rm mod}\,\, {\mathbb Z}^\genus+{\mathbf \tau} {\mathbb Z}^\genus,
\eeq
where $\genus$ is the genus of the spectral curve, and $\tau=\{\tau_{i,j}\}$ is the Riemann matrix of periods of the spectral curve.
In fact, it may happen, for some integrable systems, that the matrix $\tau$ be degenerate, so that the periods could become infinite in certain directions. In that case, the Jacobian has also non periodic directions, and is a product of some power of $\mathbb C$ times a torus.
This situation which may seem non-generic, is actually often realized for many examples of interest.
Let us ignore it for the moment, and assume that degenerate cases can be obtained as limits of the non-degenerate ones.

\figureframex{14}{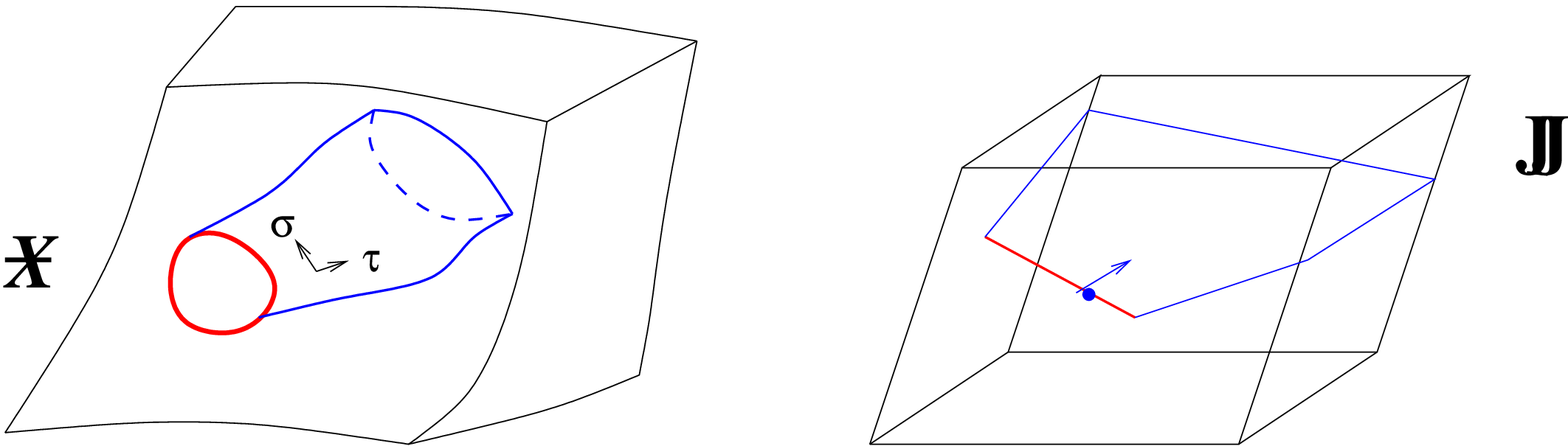}{\label{figactionangle} Under the action-angle change of coordinates, a complicated integrable motion $\vec X(t)$ in the target space $\mathfrak X$, becomes a complex time linear motion $\vec u(t)$ with constant velocity in the Jacobian $\Jac$. In other words, the worldsheet in the target space,  is a plane (a complex line) in the Jacobian.
The Jacobian is a torus, with periodicities, and thus the worldsheet can be periodic in some directions. This figure is only an "artist view" since the Jacobian should have even real dimension (never dimension 3).}

The complex time evolution of the vector  $\vec X(t)$, sweeps a surface embedded into $\CYX$, which we call a {\bf worldsheet}.
In other words, every classical solution of the equations of motion, corresponds to a worldsheet.

The 1-dimensional real curves $\vec X(\sigma+iT),\, \sigma\in \mathbb R$ at fixed $T\in\mathbb R$, are called "strings". The worldsheet is indeed the surface swept by a string as $T$ sweeps $\mathbb R$.

\smallskip

After the action-angle change of coordinates, the worldsheet in $\CYX$ is mapped to a 2-dimensional "plane" (a complex line) in the Jacobian.

The flat coordinates on the plane in the Jacobian, can be pulled back to a system of flat coordinates on the worldsheet embedded in $\CYX$ (see fig. \ref{figflcoordwsheet}).
\figureframex{14}{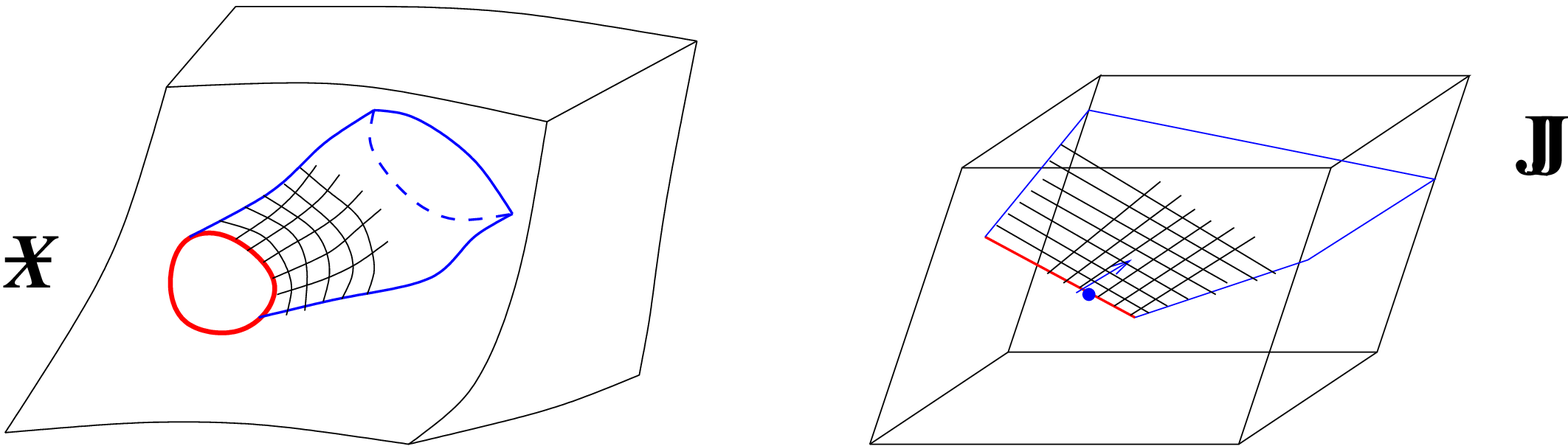}{\label{figflcoordwsheet} The flat cartesian coordinates on the plane in the Jacobian, provide a system of flat coordinates on the worldsheet.}

\medskip

\br
Another usual formulation of integrability, is the existence of some flat connection. Here, we see that the flat connection can be realized as the pullback of the parallel transport in the plane, i.e. the pullback of the trivial flat connection in the Jacobian, by the action-angle change of coordinates.
\er

\subsection{Boundaries and branes}

There is not a unique choice of flat coordinates on a plane.
Let us see one canonical choice, adapted to a choice of specific boundary conditions called Branes.

\smallskip

In what follows, we wish to enumerate worldsheets having certain topologies, and, if we want to have only finite numbers, we need to prescribe some constraints. In particular we want to ensure that there is a unique choice of local flat coordinate $t$ on each worldsheet.

\subsubsection{Branes}

For defining the boundary conditions, one first fixes a lattice vector $ \vec v \in {\mathbb Z}^\genus+{\mathbf \tau} {\mathbb Z}^\genus$. This vector, which we shall call {\bf polarization}, is a modulus of the boundary which is kept fixed from
now on. In the following, we consider only boundaries with the same polarization.

\medskip

Then, we want to consider worldsheets, whose boundaries have the  topology of a circle and such that in the Jacobian, the boundary is a straight horizontal line parallel to $\vec v$: indeed, the boundary can be a circle only if the straight line in the Jacobian is parallel to a periodic direction, i.e. to a period lattice, and we choose it to be the polarization $\vec v$ .

\medskip

\bd
{\bf D-brane:}

A worldsheet is said to have an D-brane boundary condition with polarization $\vec v$  if and only if,
in action angle coordinates, the boundary is  a straight line in the Jacobian parallel (with a real scalar factor) to the polarization $\vec v$.

\ed

\subsubsection{Cannonical choice of flat coordinate}

Our goal now, is to enumerate worldsheets having $D-brane$ boundary conditions.
In order to have only a finite number of them, we need to specify some extra conditions.

\smallskip

We consider worldsheets having D-brane boundaries with a marked point on the boundary and with a given "length" parameter $l\in \mathbb R_+$ (also called "perimeter" of the boundary).

\smallskip

Given a marked point on the boundary and a length parameter $l$,  we choose the unique cartesian coordinate $t$ on the plane in the Jacobian, by choosing the origin $t=0$ at the marked point and the unit such that $t=l$ corresponds to $\vec v$.

The boundary is then the horizontal line $\Im\, t=0$ in the plane, and it is periodic of period $t=l$.

\medskip

\br On a circle of perimeter $l$, there are $l$ possibilities to mark a point on the circle.
This means that enumerating worldsheets with a marked point on the boundary, or worldsheets without marked points, merely amounts to multiplying by $l$.
\er

\bigskip

So, a given marked point and length  provide a unique choice of time coordinate on the plane, i.e. a unique flat coordinate on the worldsheet, at least in a vicinity of the boundary.

\figureframex{14}{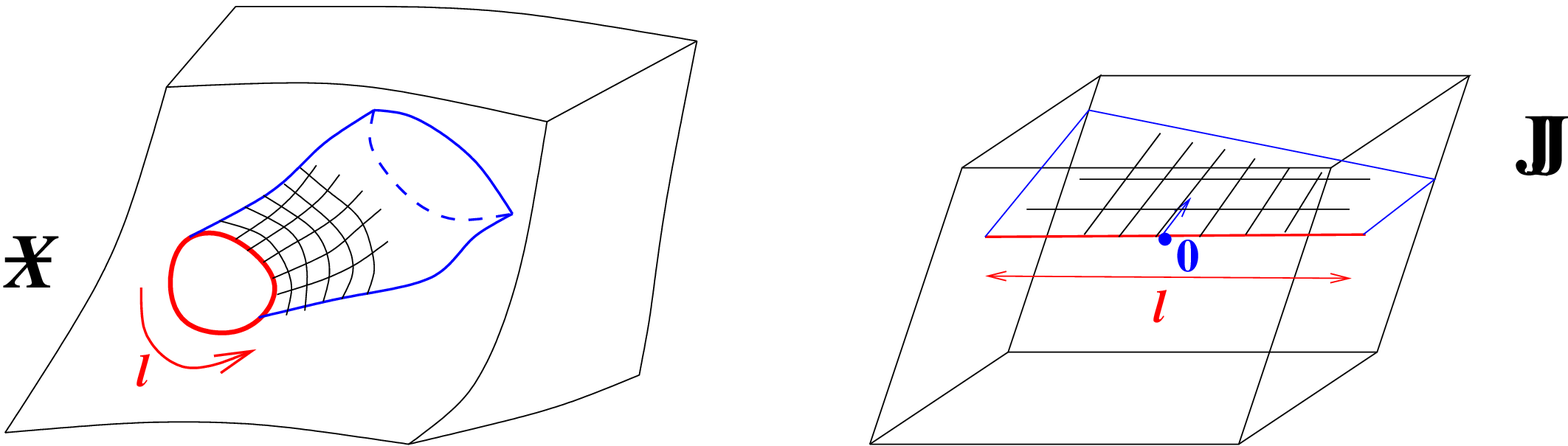}{\label{figcoordwinding}
If we consider worldsheets whose boundary is a circle which is a straight line in the Jacobian, and  with length $l$, we choose the unique coordinate $t$ on the plane in the Jacobian such that the marked point has coordinate $t=0$ and the lattice period in the Jacobian has coordinate $t=l$. That provides a unique canonical choice of flat coordinates on the worldsheet, at least in a vicinity of the boundary.}

We call the lines:

$\bullet$ $\Im\, t={\rm constant}$:  horizontal trajectories,

$\bullet$ $\Re\, t={\rm constant}$: vertical trajectories.

\smallskip
The boundary is the horizontal trajectory $\Im\, t=0$, and the worldsheet is locally near the boundary, given by the Poincarr\'e half-plane $\Im\, t\geq 0$.

\br

With the same idea, we could also consider "open" worldsheets whose boundaries are vertical trajectories. That would correspond to Von Neumann boundary conditions for our branes.
We could also enumerate open strings in that framework, but for simplicity, we don't do it in this article, and postpone it to a later work.

\er

\section{Decomposition of worldsheets}

We are now interested in "counting" (with Boltzmann weight and symmetry factor) all worldsheets, which are orientable connected Riemann surfaces, of a given genus $g$ (which, once again, has nothing to do with the genus $\genus$ of the spectral curve), and a given number $k$ of boundaries with D-brane boundary conditions of polarization $\vec v$ and with a marked point on each boundary of respective lengths $l_1,\dots,l_k$.

\medskip

Since the image of a worldsheet is the plane  of equation
$$
\vec u(t) = \vec u(0)+{t\over l}\,\vec v
$$
in the Jacobian,
counting worldsheets with a given topology, amounts to count all initial conditions $\vec u(0)\in \Jac$ compatible with the sought boundary conditions and topology.

So, moduli spaces of worldsheets can be viewed as submanifolds of the Jacobian, and are naturally endowed with some measure inherited from the Jacobian.

\bigskip

Notice, that if the initial condition (the vector $\vec u(0)$ in \eq{eqmotionactangle}) is arbitrary, it is very likely that the worldsheet will not have the right topology.
In fact, it might not be compact, it might have infinite genus.
Therefore, counting only worldsheets with finite genus is very restrictive, we are counting only a very small subset of all possible worldsheets.

This should be related to the fact that we are computing only the perturbative expansion of string theory amplitudes, and the non-perturbative part is not captured by a genus expansion.
We expect that the non-perturbative part introduced in \cite{Enonpert,EMnonpert} should take into account those infinite genus worldsheets.

\subsection{Discs}

Suppose that we want to count worldsheets, with brane boundary condition as above, having the topology of a disc, i.e. planar $g=0$ and only one boundary $k=1$.
In fact, it cannot really be a disc, because the function $T=\Im\,t$ is harmonic on the worldsheet and constant on the boundary. That would be  impossible on a simply connected domain.
This means that there must exist at least one singularity of $T=\Im\,t$ inside the disc, in other words  we have a "punctured" disc.
In some sense, this is an infinite half-cylinder, but by abuse of language, we shall continue to call it a disc.

\medskip

The punctures can sit in some critical submanifolds in the target space, for instance the non-compact directions of the target space.
Our target space and integrable theory may be such that there can exist several kinds of punctures.
We shall always need to specify which kind of punctured disc we are talking about.

\subsection{Moduli spaces}

Let us consider worldsheets of some genus $g$, with $k$ brane boundaries of given common polarization, and with respective lengths $l_1,\dots,l_k$.
The topology will be called stable if
$$
2-2g-k<0.
$$
Discs ($k=1,g=0$) and cylinders ($k=2,g=0$) are not stable.

\medskip
We define:
\bd
Let ${\cal M}_{g,k}(p_1,l_1;p_2,l_2;\dots;p_k,l_k)$ be the set of all oriented connected worldsheets of genus $g$, with $k$ brane boundaries with marked points and lengths $l_1,\dots,l_k$, sitting on branes
labeled $p_1,\dots,p_k$ respectively, and quotiented by additions of non-singular bare cylinders at the boundaries 
(see remark below).

${\cal M}_{g,k}(p_1,l_1;p_2,l_2;\dots;p_k,l_k)$ has an orientifold structure, i.e. each worldsheet is counted quotiented by its symmetries.

Let:
\beq
\NN_k^{(g)}({p_1},l_1;{p_2},l_2;\dots;{p_k},l_k) \quad ":="\quad \# {\cal M}_{g,k}(p_1,l_1;p_2,l_2;\dots;p_k,l_k),
\eeq
be the number of elements in ${\cal M}_{g,k}(p_1,l_1;p_2,l_2;\dots;p_k,l_k)$ where each worlsheet $\Sigma$ is counted with a Boltzmann weight $\ee{-{\rm action}(\Sigma)}$ coming from the action of our integrable system, and with a symmetry factor $1/\#{\rm Aut}(\Sigma)$ if it has non trivial automorphisms. More precisely:
\beq
\NN_k^{(g)}({p_1},l_1;{p_2},l_2;\dots;{p_k},l_k)\,:=\, \sum_{\Sigma\in{\cal M}_{g,k}(p_1,l_1;p_2,l_2;\dots;p_k,l_k)}\,\, {\ee{-{\rm action}(\Sigma)}\over \#{\rm Aut}(\Sigma)}.
\eeq

\ed

Notice that stable Riemann surfaces (i.e. $2-2g-k<0$) have a finite number of automorphisms.

The Boltzmann weight $\ee{-{\rm action}}$ depends on the integrable system, i.e. on the moduli of the target space, as well as some additional parameters and  the moduli of the brane boundaries.
We will not have to know the weight, and we shall prove that the topological recursion holds without having to specify the weight. This weight is encoded in the spectral curve or in the disc amplitude.

\medskip

Let us explain what we mean by  counting worldsheets, "modulo addition of non-singular cylinders at the boundary" (see fig \ref{figsurfmodcyl}).
This means that if a worldsheet is obtained from another, by analytically extending the flat coordinates near the boundary, both worldsheets are in the same equivalence class and should be counted only once.
This means that ${\cal M}_{g,k}(p_1,l_1;p_2,l_2;\dots;p_k,l_k)$ is locally independent of $p_1,\dots, p_k$. It is invariant under small changes of $p_1,\dots,p_k$.
But it can change when one of the $p_i$'s approaches a special brane, a branching or a puncture.

\figureframex{9}{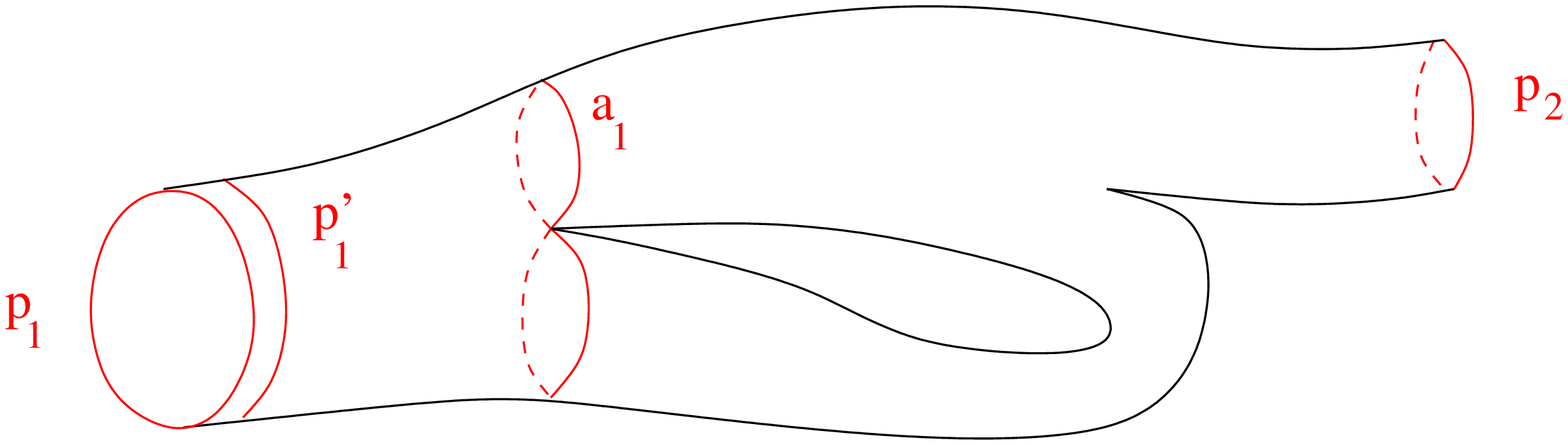}{\label{figsurfmodcyl}
Worldsheets ending on brane $p_1$ or $p'_1$ are considered equivalent, and are counted only once in ${\cal M}_{1,2}$. But they are not equivalent to worldsheets ending on $a_1$.}

\bigskip

The purpose of the next sections is to show how to compute those "weighted numbers of worldsheets" in terms of the spectral curve of the integrable system. We will see that they can be computed recursively by the topological recursion.

\subsection{Branchings}

Let us consider a worldsheet in ${\cal M}_{g,k}(p_1,l_1;p_2,l_2;\dots;p_k,l_k)$, with $2-2g-k<0$ and $k>0$.

Notice that if $2-2g-k<0$, the worldsheet is neither a cylinder nor a disc. It is locally a half-cylinder near its boundaries, but it cannot keep the topology of a cylinder under time evolution, i.e. it cannot be globally mapped to a plane bijectively.
There must be some time at which the flat coordinate becomes singular.

\medskip

Consider one of the boundaries (let us say the first one $p_1,l_1$) with its marked point and length $l_1$, and choose the unique flat coordinate $t$ defined as above in the vicinity of that boundary.

For small times $\Im\, t$, the worldsheet is a cylinder, and horizontal trajectories are circles winding  around the worldsheet.
However, since the worldsheet is not globally a cylinder, there must exist a time at which the horizontal trajectory is no longer a circle.

Several situations can occur:
\begin{itemize}

\item The horizontal trajectories can hit another boundary. This would lead to a worldsheet with the topology
of a cylinder since all boundaries are parallel in the Jacobian, that is to say $g=0$ and $k=2$ which is not considered here.

\item The horizontal trajectories can hit a puncture. This would mean that our worldsheet would be a  disc, i.e. $g=0$ and $k=1$, which is not either the situation we wish to consider here.

\item They can hit a singularity where the horizontal trajectory gets  pinched and splits into two (or more) connected components, the generic situation corresponding to two components (see fig. \ref{figsplitting}). The circle  at time $\Im t< t_c$, splits into a "figure of 8" at $t_c$.
After $t_c$, the worldsheet splits into two half-cylinders.

\item In fact, just by going backwards in time, we see that there are also singularities at which the horizontal trajectory becomes a half of a figure of 8, where another half-cylinder could join.  Again, at $t=t_c$, the horizontal trajectory is made of two circles, i.e. a "figure of $8$".

\end{itemize}

\smallskip
Let us call "branch points" $a_i$, the points in the moduli space of branes (whatever it is) at which such branchings may occur.
From now on, we assume that our integrable system be such that we have only a finite number of
branchpoints\footnote{Remark that the branch points depend on the choice of integrable system and polarization.}.

\figureframex{14}{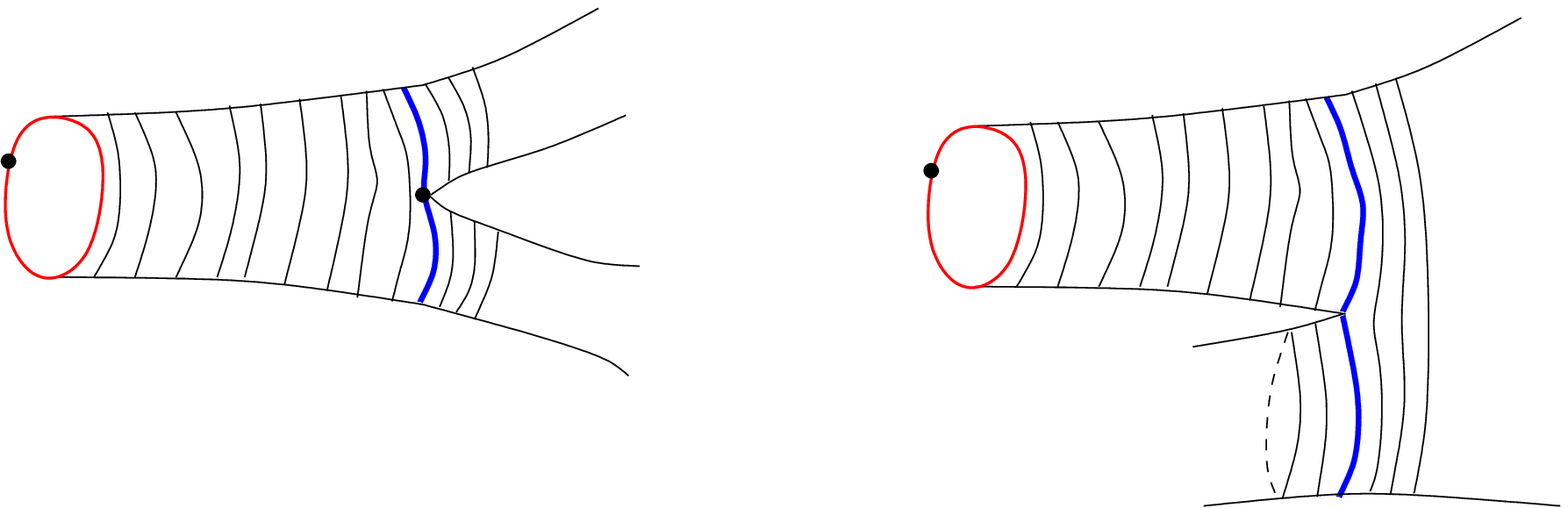}{\label{figsplitting}
The horizontal trajectory hits a singularity for the first time at time $t_c$, at which the flat coordinate is ill-defined and the circle splits into a figure of "8". Two possibilities may occur: after $t_c$ the cylinder splits into two half-cylinders, or it merges with another cylinder to make a bigger cylinder.
Remark that the second possibility is the same as the first one, under time reversal.
In both cases, the horizontal trajectory at time $\Im\,t_c$ is a figure of "8".
Situations where more than two cylinders join, or split, are non generic, and should bring a vanishing contribution to the generating function counting worldsheets.
}

\bigskip

\bd

We shall call a "bare propagator" a piece of an open worldsheet, $0<\Im\, t<\Im\, t_c$ which is a cylinder, where the flat coordinate is globally defined, and all horizontal trajectories are circles, and such that at $t=t_c$, the horizontal trajectory is a figure of 8
(see fig. \ref{figshort}).

\smallskip

We denote such a bare propagator:
\beq
\widehat{S}(p_0,l_0;a_i,l)
\eeq
where the horizontal trajectory $\Im\, t=0$ is on brane $p_0$, with length $l_0$, and the horizontal trajectory $\Im\,t=\Im\,t_c$ is on the critical brane $a_i$, with length $l=l_0$.
\ed

\figureframex{15}{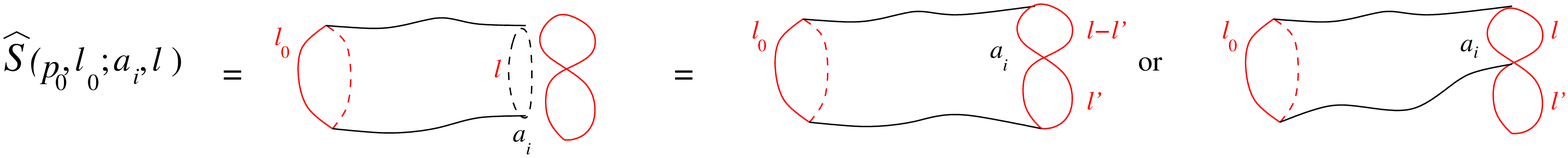}{\label{figshort} The bare propagator: its interior is a cylinder, bounded by horizontal trajectories. One side is a circle, the other side is made of two circles glued at a branchpoint.
Either the circle of length $l_0=l$ is pinched into two circles $l=l'+(l-l')$, or it is one half of a pinched circle of total size $l''=l+l'$.}

The main property, is that beyond time $t_c$, the horizontal trajectories should be conformally continued into two half cylinders.

\subsection{Recursive decomposition of the worldsheet into propagators and cylinders}\label{secbij}

Consider $g\geq 0$ and $k\geq 0$ such that $2-2g-(k+1)<0$, and
consider a worldsheet $\Sigma\in {\cal M}_{g,k+1}(p_0,l_0;p_1,l_1;p_2,l_2;\dots;p_k,l_k)$.
Consider its first boundary, of length $l_0$, with a marked point, ending on brane $p_0$, and consider the unique flat coordinate $t$ on the worldsheet near this boundary.

\figureframex{15}{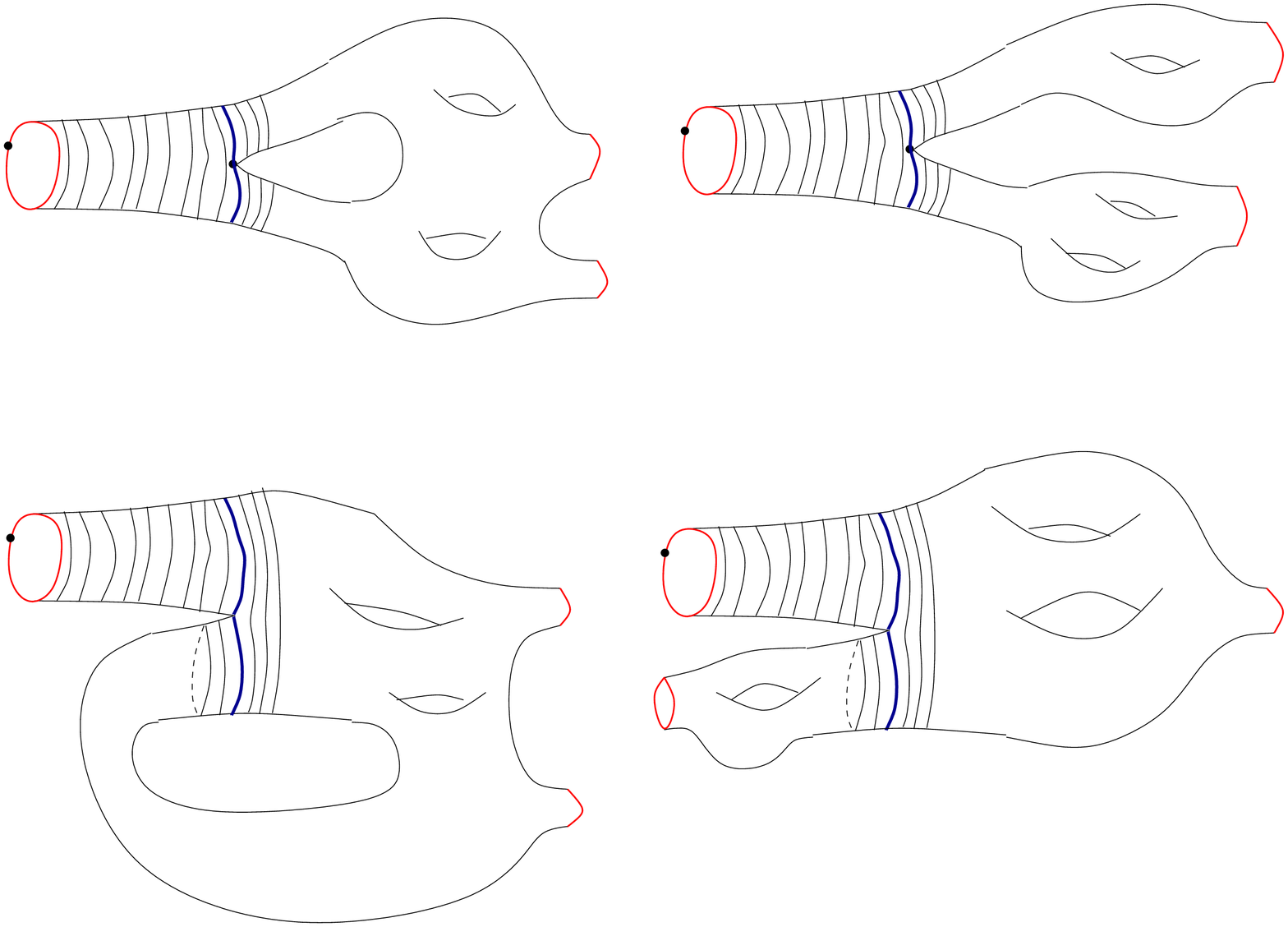}{\label{figsplitting1} Consider a worldsheet $\Sigma\in {\cal M}_{g,k+1}(p_0,l_0;p_1,l_1;\dots;p_k,l_k)$ which is not a disc or a cylinder. Start from the boundary $(p_0,l_0)$, consider the horizontal trajectories defined from that boundary. There must exist a smallest time $t_c$ at which the horizontal trajectory stops being a circle. The piece of surface $0<\Im\, t<\Im\, t_c$ is a bare propagator. If we remove the bare propagator and the figure of 8 critical trajectory from $\Sigma$, we get a worldsheet $\Sigma'$. $\Sigma'$ is either connected or disconnected.}

Since the worldsheet doesn't have globally the topology of a disc or cylinder, there must exist a smallest time $t_c$, $\Im\, t_c>0$, at which the flat coordinate becomes ill defined and a branching occurs.
In other words, there exists a, generically unique, time $t_c$ at which we reach some  branchpoint $a_i$, and therefore the worldsheet $\Sigma$ contains a bare propagator $\widehat{S}(p_0,l_0;a_i,l)$. We emphasize  that this propagator is uniquely defined.

Let us call $\Sigma'$ the worldsheet obtained by removing the bare propagator from $\Sigma$:
\beq
\Sigma' = \Sigma \setminus \widehat{S}(p_0,l_0;a_i,l).
\eeq

$\Sigma'$ has again brane boundary conditions (boundaries are indeed  horizontal trajectories parallel to $\vec v$), it has $k+2$ boundaries since one of the boundaries is split,  and $\Sigma'$ is either connected or disconnected:

$\bullet$
If $\Sigma'$ is connected, it is clear that it belongs to either ${\cal M}_{g-1,k+2}(a_i,l';a_i,l-l';p_1,l_1;p_2,l_2;\dots;p_k,l_k)$ or ${\cal M}_{g-1,k+2}(a_i,l';a_i,l+l';p_1,l_1;p_2,l_2;\dots;p_k,l_k)$, i.e. in both cases to ${\cal M}_{g-1,k+2}(a_i,|l'|;a_i,|l-l'|;p_1,l_1;p_2,l_2;\dots;p_k,l_k)$ for some $l'\in \mathbb R$.

$\bullet$
If $\Sigma'=\Sigma'_+ \cup \Sigma'_-$ is disconnected, the two connected parts belong to ${\cal M}_{h,1+\# I}(a_i,|l'|;I)$ and ${\cal M}_{h',1+\#I'}(a_i,|l-l'|;I')$ for some $h,h',I,I'$ such that $h+h'=g$ and $I\uplus I'=\{p_1,l_1;\dots;p_k,l_k\}$, and $l'\in \mathbb R$.

\medskip

When it is disconnected, it may happen that one of the two connected components, let us say $\Sigma'_+$ is a punctured disc $\in {\cal M}_{0,1}(a_i,|l-l'|)$, and the other connected component $\Sigma'_-$ then belongs to ${\cal M}_{g,k+1}(a_i,|l'|;p_1,l_1;p_2,l_2;\dots;p_k,l_k)$, like $\Sigma$ itself (in particular, the other connected component can't be a disc).
In that case we may redo the same thing on $\Sigma'_-$: start from the boundary, until we reach a branchpoint, and remove the corresponding bare propagator.
We can do that recursively, until none of the connected components is a disc.

\medskip

It is thus more convenient to define a "renormalized propagator" which may include an arbitrary number of discs glued. It is defined by the property (see fig. \ref{figpropagren})
\bea\label{eqshortren}
&& S(p_0,l_0;a_i,l)  \cr
&=& \delta(l_0-l)\,\widehat S(p_0,l_0;a_i,l) \cr
&&  + 2\sum_j \int_{-\infty}^\infty d{l'}\, \widehat S(p_0,l_0;a_j,l_0) \cup S(a_j,|l'|;a_i,l) \cup {\rm Disc}(a_j,|l_0-l'|). \cr
\eea

\figureframex{12}{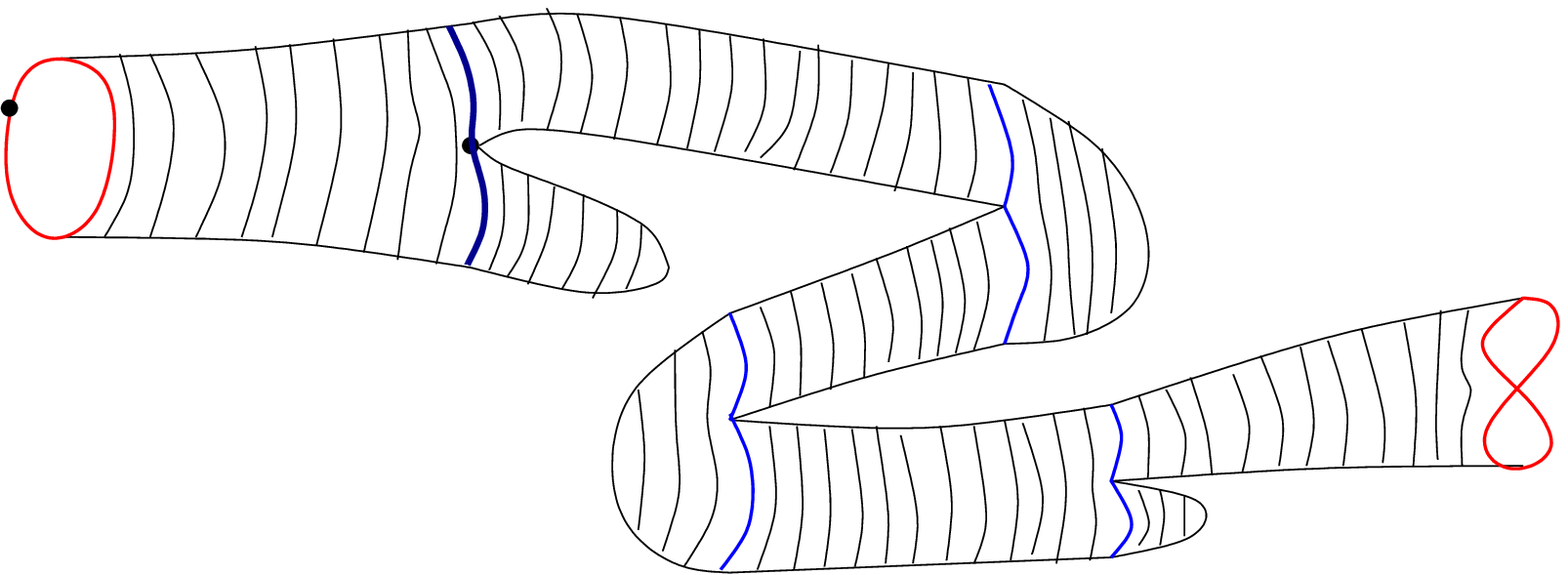}{\label{figpropagren} The renormalized propagator, is obtained by following the horizontal trajectories from the first boundary. Each time a critical trajectory is met, the surface may split into two disconnected parts. We recursively do that until none of the connected components is a disc. The renormalized propagator thus contains a certain number of bare propagators and discs. It ends at a critical trajectory.}

\bigskip

Removing the renormalized propagator from a worldsheet $\Sigma$, means removing a propagator from the first boundary, and if one of the connected components is a disc, then remove again a propagator from the other connected component until none of the connected components is a disc:
\beq
\Sigma'' = \Sigma \setminus S(p_0,l_0;a_i,l).
\eeq
Notice that, since none of the connected components of $\Sigma''$ is a disc, then, each connected component of $\Sigma''$ has a Euler characteristics strictly larger than that of $\Sigma$, and therefore after repeating this procedure a finite number of times, we arrive to only propagators and cylinders.
 This is a topological recursion.

Therefore we can decompose any worldsheet $\Sigma$, in a generically unique way, into a finite number of renormalized propagators, and cylinders.

Finally we have the following (orientifold) bijection between moduli spaces:
\bea\label{eqrecursionMgk}
&&{\cal M}_{g,k+1}(p_0,l_0;p_1,l_1;p_2,l_2;\dots;p_k,l_k)\cr
&\simeq&  \sum_i\, \int_0^\infty\, dl\,\int_{-\infty}^\infty dl'\, {\cal M}_{S}(p_0,l_0;a_i,l) \times \Big[{\cal M}_{g-1,k+2}(a_i,|l'|;a_i,|l-l'|;p_1,l_1;p_2,l_2;\dots;p_k,l_k) \cr
&& {\displaystyle \cup'_{h+h'=g;I\uplus I'=\{p_1,l_1;\dots;p_k,l_k\} }} \,\, {\cal M}_{h,1+\#I}(a_i,|l'|;I)\times {\cal M}_{h',1+\#I'}(a_i,|l-l'|;I')\Big]
\eea
where $\cup'$ means that we exclude discs i.e. $(h,I)=(0,\emptyset)$ and $(h',I')=(0,\emptyset)$, and
${\cal M}_{S}(p_0,l_0;a_i,l)$ is the moduli space  of all renormalized propagators with one boundary of length $l_0$ on the brane $p_0$,
and the other boundary on brane $a_i$, of length $l$.

\medskip
This recursive decomposition of moduli spaces of worldsheets is very similar to the recursive structure of the topological recursion \eq{eqrectopstrft}.
In the next sections, we shall show that the generating functions of the volume of these moduli spaces 
indeed satisfy the topological recursion.

\subsection{Skeleton graph of the worldsheet}\label{secgraphs}

Another consequence of that decomposition, is that we can associate a graph to any worldsheet.

Indeed, chose the first boundary and remove renormalized propagators, i.e. remove propagators until none of the two connected components is a disc.
Then, draw the splitting horizontal trajectory, that is the boundaries of the renormalized propagators, on the worldsheet, and proceed recursively, until it remains only cylinders.

We have thus drawn some dividing circles on each world sheet.
On each renormalized propagator, let us draw an arrowed line from the marked point on the circle boundary, to the branchpoint. On each cylinder, let us draw an unoriented line between the 2 marked points.

$$
{\mbox{\epsfxsize=4.truecm\epsfbox{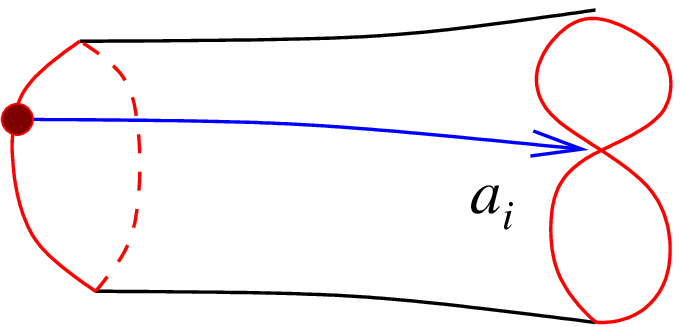}}}
\qquad
{\mbox{\epsfxsize=4.truecm\epsfbox{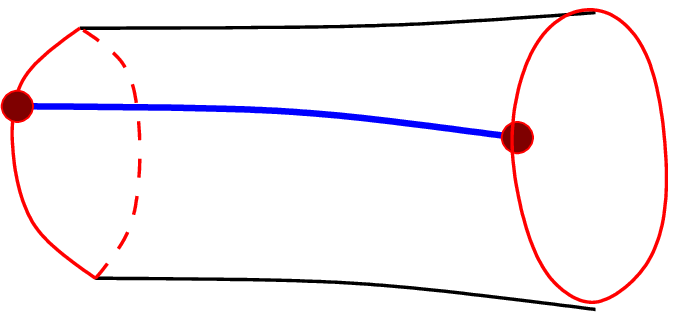}}}
$$

We obtain a graph drawn on each worldsheet, whose vertices are labeled by branchpoints.
These are exactly the graphs of \cite{EOFg}.
Those graphs have $2g-2+k$ vertices, $2g-2+k$ arrowed edges, $k-1$ external non-arrowed external edges, $g$ internal non-arrowed edges, forming $g$ loops, and such that the $2g-2+k$ arrowed lines form a tree rooted at the first boundary and going through all vertices. The edging is also constrained by the following rule:
non-arrowed lines can only connect two vertices if one is the descendent of the other along the arrowed tree,
see \cite{eynloop1mat, EOFg}.

\figureframex{14}{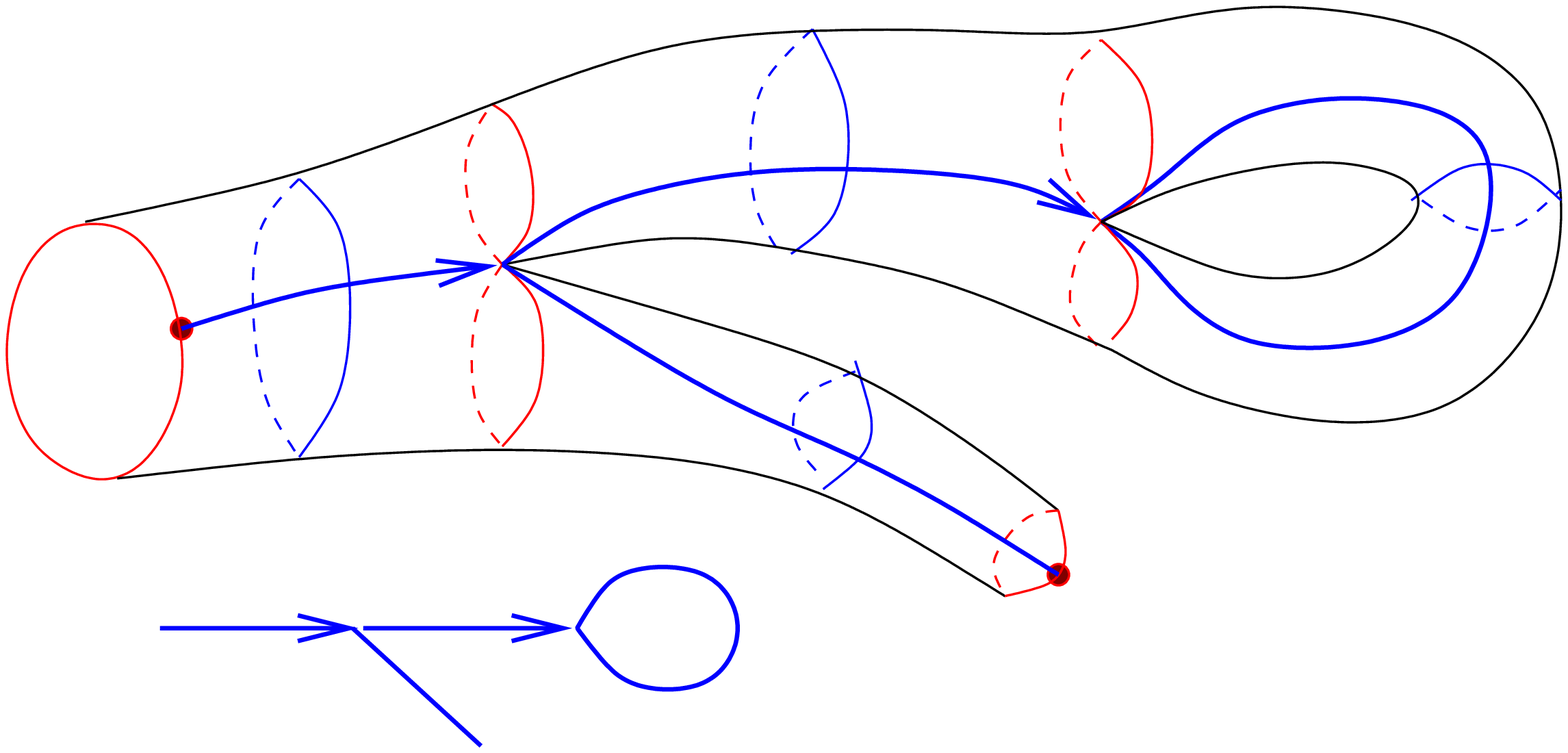}{\label{figsurf21graph} Example: the graph on a worldsheet of ${\cal M}_{1,2}$. Each worldsheet has a unique graph, once we have chosen an entrance boundary. Notice that the dual (blue circles in the middle of each cylinder), gives a canonical pant decomposition of the worldsheet.}

\br
The graph obtained depends on a choice of a "first boundary", another choice could lead to another graph.
\er

One can see that the dual of the graph (see fig. \ref{figsurf21graph}), obtained by drawing circles dividing every cylinder into two half-cylinders, provides a pant decomposition of the worldsheet.
But contrarily to the usual Teichm\"uller spaces approach, here, thanks to our integrable system, we have for each worldsheet and choice of "first boundary", a unique pant decomposition.
We don't have to consider a quotient by the mapping class group. In some sense we have already chosen one canonical representant in each class.

\br
Different worldsheets in ${\cal M}_{g,k}$ can have different graphs. This allows to decompose the moduli space ${\cal M}_{g,k}$ into a finite number of cells labeled by graphs.
\figureframex{8}{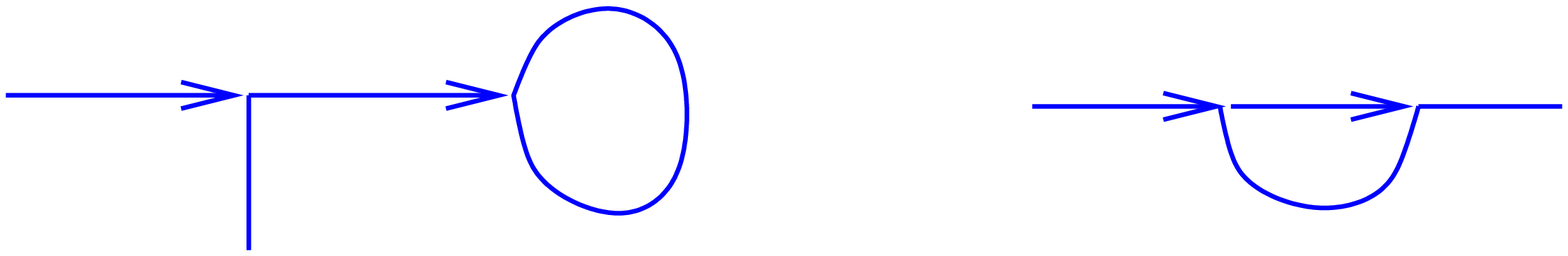}{The two possible graphs obtained for worldsheets in ${\cal M}_{1,2}$.}
Somehow this is in the same spirit as the Strebel-foliation used by Kontsevich, or the Teichm\"uller pant decomposition.

\er

\section{Topological recursion for the amplitudes}

In this section, using standard methods of combinatorics, and in particular Laplace transforms, we translate the recursive decomposition of our moduli spaces into relations for the generating functions $\NN_k^{(g)}({p_1},l_1;{p_2},l_2;\dots;{p_k},l_k)$.

In the preceding section, we showed that integrability allows to get a unique foliation of every worldsheet through the
use of a flat connection built from the integrability of the considered system. This allows to get a
cell decomposition of the moduli space of surfaces labeled by the graphs of \cite{eynloop1mat, EOFg} described in section \ref{secgraphs}. Moreover,
this gives a bijective procedure to build every worldsheet by gluing discs and cylinders. In this section, we
translate this bijection into a recursive relation among the amplitudes $\NN_k^{(g)}({p_1},l_1;{p_2},l_2;\dots;{p_k},l_k)$ which are the generating functions counting the elements of ${\cal M}_{g,k}$ (with Boltzmann weights and symmetry factors).

\subsection{Disc amplitude}

For the moment, we shall not explain how to compute the disc amplitudes
$
\NN_1^{(0)}(p_1,l_1).
$
We shall assume those numbers to be given.
The way they depend on the integrable system, or how they are related among themselves, will be explained later in section \ref{secreconstructspcurvedisc}. Let us just mention that they are "difficult" to compute.

\smallskip

We also emphasize that what we call a disc is really a "renormalized disc", i.e. it may contain several punctures and branchings, the flat coordinate needs not be globally well defined on it. Renormalized discs can be obtained by gluing propagators and bare discs recursively, in the same way we defined the renormalized propagator.
But since we don't know yet the generating function of bare discs, we shall not perform that construction here.

\smallskip

For an $a_i$ brane, we define the generating function as a function of a formal complex variable $z\in \mathbb C$, by a Laplace transform:
\beq
\widetilde W_1^{(0)}(a_i,z) = \int_0^\infty dl\,\, \ee{-zl}\,\, \NN_1^{(0)}(a_i,l).
\eeq

Notice that $\NN_1^{(0)}(a_i,l)$ counts discs with a marked point, and therefore:
\beq
\NN_1^{(0)}(a_i,l) = l\,\, \widehat \NN_1^{(0)}(a_i,l)
\eeq
where $\widehat \NN_1^{(0)}(a_i,l)$ counts discs without marked points.
This implies that $\widetilde W_1^{(0)}(a_i,z)$ is a derivative with respect to $z$, and this is why $\widetilde W_1^{(0)}(a_i,z)\,dz$ should be thought of as a differential form.

\smallskip

From now on, we shall assume that our integrable system is "regular", i.e. such that $\widetilde W_1^{(0)}(a_i,z)$ is analytical near $z=0$, and that $\widetilde W_1^{(0)}(a_i,z)+\widetilde W_1^{(0)}(a_i,-z)$ has only a double zero, and not a higher order  zero.
We say that our integrable system is critical when we have higher order zeroes.

\subsection{Propagator amplitude}

The bare propagator means the non-renormalized propagator, i.e. a piece of a worldsheet, where the flat coordinate is globally well defined.
Let us denote its amplitude by:
$
\ovl K(p_0,l_0;a_i,l)
$.
Because there is no singularity, the lengths are conserved, and thus the amplitude can be non zero only if $l_0=l$.

\medskip

The renormalized propagator is obtained by gluing discs to one of the 2 boundaries, i.e. its amplitude
$
K(p_0,l_0;a_i,l)
$
satisfies the relation (see fig.\ref{figpropagren}):
\bea
&& K(p_0,l_0;a_i,l) \cr
&=& \delta(l-l_0)\,\ovl K(p_0,l_0;a_i,l) \cr
&& + 2 \sum_j \int_{-\infty}^\infty\, dl'\,\,
\ovl K(p_0,l_0;a_j,l_0)\,\,
\NN_1^{(0)}(a_j,|l_0-l'|)\,\,
K(a_j,|l'|;a_i,l) .\cr
\eea

This relation doesn't determine $K$ or $\ovl K$, and we shall see later how to determine them. However, it allows to
express $K$ in terms of $\ovl K$ and $\NN_1^{(0)}$.

\medskip
In Laplace transform we define the generating function:
\beq
K_{p_0,l_0}(a_i,z)  = \int_0^\infty\, dl\,\, \ee{-z\, l}\,\, K_{p_0,l_0;a_i,l} \, .
\eeq

\subsection{Annulus and propagator amplitude}

We call annulus or cylinder the elements of ${\cal M}_{0,2}$.

\medskip

Let us consider a bare cylinder, i.e. a worldsheet starting on a brane $p_1$ with length $l_1$, and ending on $p_2$ with length $l_2$.
"Bare" means that the flat coordinate is globally well defined on the cylinder and doesn't encounter any singularity between the two boundaries.

It is obvious then, that all horizontal trajectories have the same length and thus $l_1=l_2$, i.e.
$$
\ovl\NN^{(0)}_2(p_1,l_1;p_2,l_2) =0 \,\qquad {\rm if} \,\,l_1\neq l_2
$$
for any $p_1,p_2$.
Also, since we count worldsheets modulo gluing a bare cylinder, we see that if the boundaries $p_1$ and $p_2$ are close enough, all cylinders are equivalent.

If $p_1$ and $p_2$ are close enough, it is clear that there is only one possible cylinder, and there is $l$ possibilities of choosing a marked point on the second boundary, so that:
$$
\ovl\NN^{(0)}_2(p_1,l;p_2,l) = l.
$$

Then, consider the renormalized cylinders.
This means that between branes $p_1$ and $p_2$, the flat coordinate $t$ may encounter branchings where one branch ends on a disc.
In other words we may glue many discs recursively.
The process of gluing discs at branching is already included in the renormalized propagator (see fig \ref{figKBW}), so that we have:
\bea\label{eqsylbarren}
\NN^{(0)}_2(p_1,l_1;p_2,l_2)
&=& \delta(l_1-l_2)\,\ovl \NN^{(0)}_2(p_1,l_1;p_2,l_1) \cr
&& + 2\sum_j \int_{l_2}^\infty\,dl'\,\, K(p_1,l_1;a_j,l'-l_2)\,\, \NN^{(0)}_1(a_j,l')\,\, \ovl \NN^{(0)}_2(a_j,l_2;p_2,l_2) \cr
&& + 2\sum_j \int_{0}^{l_2}\,dl'\,\, K(p_1,l_1;a_j,l_2-l')\,\, \NN^{(0)}_1(a_j,l')\,\, \ovl \NN^{(0)}_2(a_j,l_2;p_2,l_2) \cr
&& + 2\sum_j \int_{0}^\infty\,dl'\,\, K(p_1,l_1;a_j,l'+l_2)\,\, \NN^{(0)}_1(a_j,l')\,\, \ovl \NN^{(0)}_2(a_j,l_2;p_2,l_2) \cr
&=& \delta(l_1-l_2)\,\ovl \NN^{(0)}_2(p_1,l_1;p_2,l_1) \cr
&& + 2\sum_j \int_{-\infty}^\infty\,dl'\,\, K(p_1,l_1;a_j,|l_2-l'|)\,\, \NN^{(0)}_1(a_j,|l'|)\,\, \ovl \NN^{(0)}_2(a_j,l_2;p_2,l_2). \cr
\eea

If the second brane $p_2=a_i$ is at a branchpoint, this means the cylinder ends exactly where the flat coordinate degenerates. The cylinder is thus exactly the same as the propagator, except that we need to glue a disc at the other half-boundary, so that the boundary has the topology of a circle. That translates to:
\bea\label{eqcylshort2}
{1\over l_2}\,\NN^{(0)}_2(p_1,l_1;a_i,l_2)
&=&   \int_{-\infty}^\infty\,dl'\,\, \,K(p_1,l_1;a_i,|l_2-l'|)\,\, \NN^{(0)}_1(a_i,|l'|)\,
\eea
Notice that we have a $1/l_2$ in the left hand side, and a $1/2$ (in fact $2$ times $1/2$) in the right hand side, because of symmetry factors of marking a point on the boundary.

\figureframex{13}{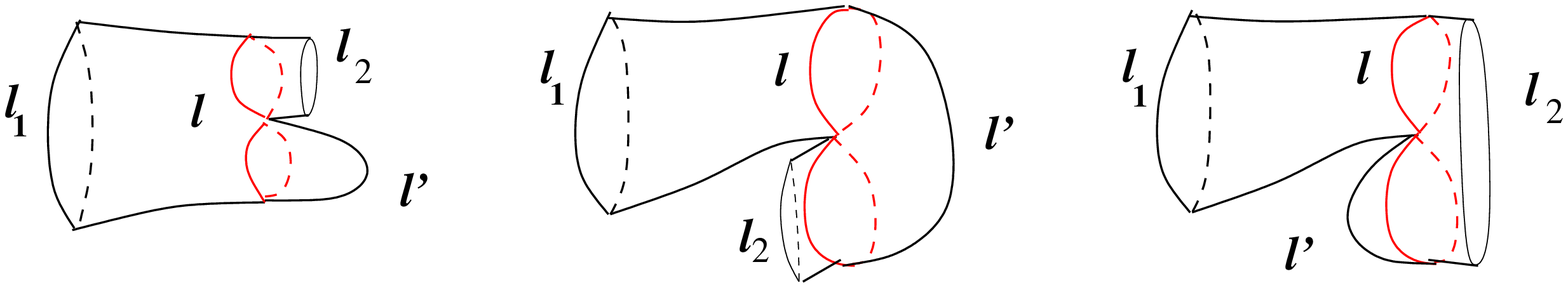}{The 3 terms in \eq{eqsylbarren} .\label{figKBW}}




Notice that this last relationship, together with \eq{eqsylbarren} implies:
\begin{eqnarray*}
\NN^{(0)}_2(p_1,l_1;p_2,l_2)
&=& \delta(l_1-l_2)\,\ovl \NN^{(0)}_2(p_1,l_1;p_2,l_1) \cr
&& + \sum_j {1\over l_2}\,\,\NN^{(0)}_2(p_1,l_1;a_j,l_2)\,\, \ovl \NN^{(0)}_2(a_j,l_2;p_2,l_2). \cr
\end{eqnarray*}
which just means that gluing two cylinders gives a cylinder\footnote{Since we consider marked points on the boundary, we need to divide by $l_2$ in order to forget the marking on the intermediate boundary.}. This is the self-reproducing property of cylinders.

\medskip
We define the generating functions:
\beq
B(p_0,n_0; a_i,z)  = \,\,\int_{0}^\infty\,dl\,\, \ee{-zl}\,\, \NN^{(0)}_2(p_0,l_0;a_i,l)\,
\eeq

\eq{eqcylshort2} translates into
\bea
&& -\int_0^z\, B(p_0,l_0;a_i,z')\,dz'  \cr
&=& \int_0^\infty\,\, {dl_2\over l_2}\,\ee{-z l_2}\,\, \NN_2^{(0)}(p_0,l_0;a_i,l_2) \cr
&=&  \,\int_0^\infty\,\, dl\,\int_0^{l}\,\, dl' \,\ee{-z (l-l')}\,\, K(p_0,l_0;a_i,l)\, \NN_1^{(0)}(a_i,l') \cr
&& + \,\int_0^\infty\,\, dl\,\int_l^{\infty}\,\, dl' \,\ee{+z (l'-l)}\,\, K(p_0,l_0;a_i,l)\, \NN_1^{(0)}(a_i,l') \cr
&& + \,\int_0^\infty\,\, dl\,\int_0^{\infty}\,\, dl' \,\ee{-z (l'+l)}\,\, K(p_0,l_0;a_i,l)\, \NN_1^{(0)}(a_i,l') \cr
&=&  \,\int_0^\infty\,\, dl\,\int_0^{\infty}\,\, dl' \,\ee{-zl}(\ee{zl'}+\ee{-zl'})\,\, K(p_0,l_0;a_i,l)\, \NN_1^{(0)}(a_i,l') \cr
&=& K(p_0,l_0;a_i,z)\,\,(\widetilde W_1^{(0)}(a_i,z)+\widetilde W_1^{(0)}(a_i,-z))
\eea

i.e.
\beq
K(p_0,l_0;a_i,z) = -{\int_{z'=0}^z\, B(p_0,l_0;a_i,z')\,dz'\over \,(\widetilde W_1^{(0)}(a_i,z)+ \widetilde W_1^{(0)}(a_i,-z))}
\eeq
This relationship is merely the combinatoric relation illustrated by fig \ref{figKBW}, and is to be compared with \eq{eqdefK}.

\subsection{Topological recursion}

The bijective procedure of section \ref{secbij} tells us that moduli spaces of stable topologies can be decomposed recursively.
The bijection \eq{eqrecursionMgk} clearly translates into the following relation among amplitudes:
\bea
&& \NN^{(g)}_{k+1}(p_0,l_0;p_1,l_1;\dots;p_k,l_k) \cr
&=& \sum_i \int_0^{\infty} dl \int_{-\infty}^\infty dl'\,  K(p_0,l_0;a_i,|l+l'|)  \Big[ \NN^{(g-1)}_{k+2}(a_i,l;a_i,|l'|;p_1,l_1;p_2,l_2;\dots;p_k,l_k) \cr
&&  \qquad \qquad \quad+ {\displaystyle \sum'_{h+h'=g;I\uplus I'=\{p_1,l_1;\dots;p_k,l_k\} }}\,\, \,\, \NN^{(h)}_{1+\#I}(a_i,l;I)\,\,\NN^{(h')}_{1+\#I'}(a_i,|l'|;I') \Big] \cr
&=& \sum_i {1\over 2i\pi}\int_{i\mathbb R-0} dz
\quad \int_0^{\infty}\, dl'' \int_{-\infty}^\infty\, dl\int_{-\infty}^\infty\, dl'\,\, \ee{-z(l''+l+l')} \cr
&& K(p_0,l_0;a_i,l'')  \Big[ \NN^{(g-1)}_{k+2}(a_i,|l|;a_i,|l'|;p_1,l_1;p_2,l_2;\dots;p_k,l_k) \cr
&&  \qquad \qquad \quad+ {\displaystyle \sum'_{h+h'=g;I\uplus I'=\{p_1,l_1;\dots;p_k,l_k\} }}\,\, \,\, \NN^{(h)}_{1+\#I}(a_i,|l|;I)\,\,\NN^{(h')}_{1+\#I'}(a_i,|l'|;I') \Big] \cr
\eea
where  $\sum'$ means that we exclude $(h,I)=(0,\emptyset)$ and $(h',I')=(0,\emptyset)$.

With the Laplace transforms generating functions
\beq
\widetilde W^{(g)}_{k}(a_{i_1},z_1;\dots;a_{i_k},z_k) = \int_{0}^\infty\dots\int_0^\infty dl_1\dots dl_k\,\,\, \NN^{(g)}_k(a_{i_1},l_1;\dots;a_{i_k},l_k)\,\, \prod_{j=1}^k \,\, \ee{-z_j\, l_j},
\eeq
the recursion relation can be rewritten:
\bea
&& \widetilde W^{(g)}_{k+1}(a_{i_0},z_0;a_{i_1},z_1;\dots;a_{i_k},z_k) \cr
&=& \sum_i {1\over 2i\pi}\int_{i\mathbb R-0} dz\,K(a_{i_0},z_0;a_i,z)\,\, \Big[ \widetilde W^{(g-1)}_{k+2}(a_i,z;a_i,-z;a_{i_1},z_1;\dots;a_{i_k},z_k) \cr
&& +{\displaystyle \sum'_{h+h'=g;I\uplus I'=\{a_{i_1},z_1;\dots;a_{i_k},z_k\} }}\,\,\, \widetilde W^{(h)}_{1+\# I}(a_i,z;I) \, \widetilde W^{(h')}_{1+\# I'}(a_i,-z;I') \Big].
\eea

By an easy recursion, one sees that the integration contour over $z$ can be deformed into a circle surrounding the pole at $z=0$, i.e.
\bea
&& \widetilde W^{(g)}_{k+1}(a_{i_0},z_0;a_{i_1},z_1;\dots;a_{i_k},z_k) \cr
&=& \sum_i \Res_{z\to 0} dz\,K(a_{i_0},z_0;a_i,z)\,\, \Big[ \widetilde W^{(g-1)}_{k+2}(a_i,z;a_i,-z;a_{i_1},z_1;\dots;a_{i_k},z_k) \cr
&& +{\displaystyle \sum'_{h+h'=g;I\uplus I'=\{a_{i_1},z_1;\dots;a_{i_k},z_k\} }}\,\,\, \widetilde W^{(h)}_{1+\# I}(a_i,z;I) \, \widetilde W^{(h')}_{1+\# I'}(a_i,-z;I') \Big].
\eea
This is the topological recursion written in terms of a local coordinate $z$ near each branchpoint.

\medskip
The only thing we need to do, in order to fully recover the topological recursion of \cite{EOFg}, is rewrite all those residues formula in terms of intrinsic variables on the spectral curve, rather than local coordinates.

\subsection{Closing a boundary}

Consider $2-2g-n<0$ and a worldsheet in ${\cal M}_{g,n}(p_1,l_1;\dots;p_n,l_n)$.
As we have seen, it is obtained by gluing along critical horizontal trajectories $2g-2+n$ renormalized propagators, and $g+n-1$ renormalized cylinders.

Remember that a renormalized propagator $S(p_1,l_1;a_i,l)$ is a cylinder whose second boundary ends on a critical horizontal trajectory. The first boundary ends on brane $p_1$.

However, in the topological recursion, we need to consider also renormalized propagators, whose both ends are on critical trajectories $S(a_j,l';a_i,l)$. In order for the first boundary to have the topology of a circle so that we can glue it to another propagator, we need to glue a disc on the second connected component of the first boundary.

This means that every internal renormalized propagator must contain a disc.
Notice that for the cylinders, we have the possibility of having the bare cylinder, which contains no disc.
This means that generically, a worldsheet of  ${\cal M}_{g,n}$ contains $2g+n-3$ discs.
\smallskip

If the boundaries $p_1,\dots, p_n$ are themselves critical branes $a_{i_1},\dots,a_{i_n}$, the first propagator starting on $a_{i_1}$ must also contain a disc, and therefore,
a worldsheet of  ${\cal M}_{g,n}(a_{i_1},l_1;\dots;a_{i_n},l_n)$ contains $2g+n-2$ discs.
\bigskip

\figureframex{14}{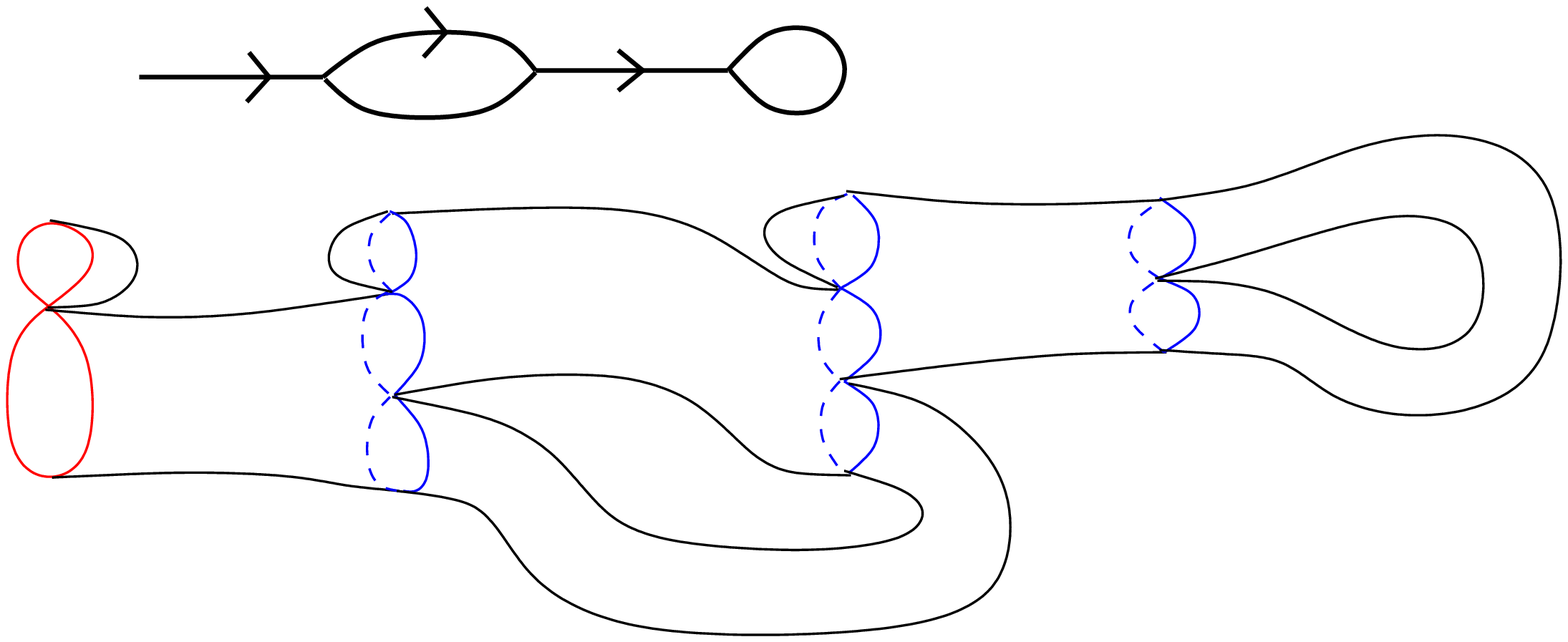}{\label{figDiscsW21} Example of a worldsheet in  ${\cal M}_{2,1}$. It can be decomposed into 3 propagators and 2 cylinders. Since the initial boundary of each propagator is a critical trajectory, we need to close the second component of the "8" by a disc. In other words the worldsheet must contain 3 discs. By cutting out one disc, we get a worldsheet in ${\cal M}_{2,2}$.}

Consider a worldsheet  $\Sigma\in{\cal M}_{g,n}(a_{i_1},l_1;\dots;a_{i_n},l_n)$ (see fig.\ref{figDiscsW21}).
Choose one among the $2g+n-2$ critical trajectories at the initial end of an internal propagator, and cut the worldsheet along that trajectory.
It gives
\beq
\Sigma  = \Sigma'\,\cup \, {\rm Disc}(a_i,l).
\qquad, \quad \Sigma'\in {\cal M}_{g,n+1}(a_{i_1},l_1;\dots;a_{i_n},l_n;a_i,l).
\eeq
Since there are $2g+n-2$ ways of doing that, we have a $2g+n-2\to 1$ application from $ {\cal M}_{g,n+1}\to  {\cal M}_{g,n}$.
This implies:
\beq
(2g+n-2)\, \NN_n^{(g)}(a_{i_1},l_1;\dots;a_{i_n},l_n) = \sum_i \int_0^\infty\, {dl\over l}\,\, \NN_{n+1}^{(g)}(a_{i_1},l_1;\dots;a_{i_n},l_n;a_i,l)\,\, \NN^{(0)}_1(a_i,l)
\eeq
(we need to divide by $l$ because otherwise the marked point is marked twice, once in $\NN_{n+1}^{(g)}$ and once in $\NN_1^{(0)}$).

In Laplace transform that gives:
\beq\label{eqrecap}
(2g+n-2)\, \widetilde W^{(g)}_n(a_{i_1},z_1;\dots;a_{i_n},z_n) = \sum_i \Res_{z\to 0}\,\, \widetilde W^{(g)}_{n+1}(a_{i_1},z_1;\dots;a_{i_n},z_n;a_i,z)\,\, \widetilde\Phi(a_i,z)\, dz
\eeq
where
\beq
d\widetilde\Phi(a_i,z)/dz = \,\widetilde W^{(0)}_1(a_i,z).
\eeq
This  relationship was also derived in \cite{EOFg} as a consequence of the topological recursion.

\subsection{Closed surfaces}

Worldsheets belonging to ${\cal M}_{g,0}$ have no boundary.
However, we shall assume that they are defined also with respect to the same polarization $\vec v$.
This means, that any worldsheet $\Sigma\in {\cal M}_{g,0}$, is also a plane parallel to $\vec v$ in the Jacobian.
We may choose any point on $\Sigma$, and draw the horizontal trajectory going through it.

So, let us choose a point on $\Sigma$, and let us  cut the worldsheet along the horizontal trajectory through that point.
The resulting worldsheet $\Sigma'$ maybe disconnected or not. It has two boundaries with  brane boundary conditions, and thus $\Sigma'$ belongs either to
\beq
{\cal M}_{g-1,2}
\qquad {\rm or}\qquad
{\cal M}_{g',1}\times {\cal M}_{g-g',1}.
\eeq
The results of previous section imply that $\Sigma'$ can  be decomposed into propagators and cylinders, and must have $2g-2$ discs (we assume $g\geq 2$).

\smallskip
Let us choose one of the discs of $\Sigma$, and choose a point on the boundary of that disc.
Let us then redefine $\Sigma'$ as $\Sigma$ cut along the horizontal trajectory going through that point.

In that case we have
\beq
\Sigma'\in {\cal M}_{0,1}(a_i,l)\times {\cal M}_{g,1}(a_i,l).
\eeq
In other words, every worldsheet $\Sigma\in {\cal M}_{g,0}$ can be obtained by gluing a disc to the boundary of a worldsheet $\in {\cal M}_{g,1}$, which can be decomposed into $2g-1$ propagators and $g$ cylinders.

Therefore every worldsheet $\Sigma\in {\cal M}_{g,0}$ can be decomposed into a disc, $2g-1$ propagators, and $g$ cylinders.

\medskip

This decomposition is not unique, it can be done for any of the $2g-2$ discs, this means that the decomposition is not bijective, but is $2g-2\to 1$.

We thus obtain:
\beq
\widetilde W_0^{(g)}=F_g = {1\over 2-2g} \sum_i \Res_{z\to a_i}\,\, \widetilde W_1^{(g)}(a_i,z)\,\, \widetilde\Phi(a_i,z)\, dz.
\eeq

This is precisely how $F_g$'s are defined in \cite{EOFg}.

\figureframex{14}{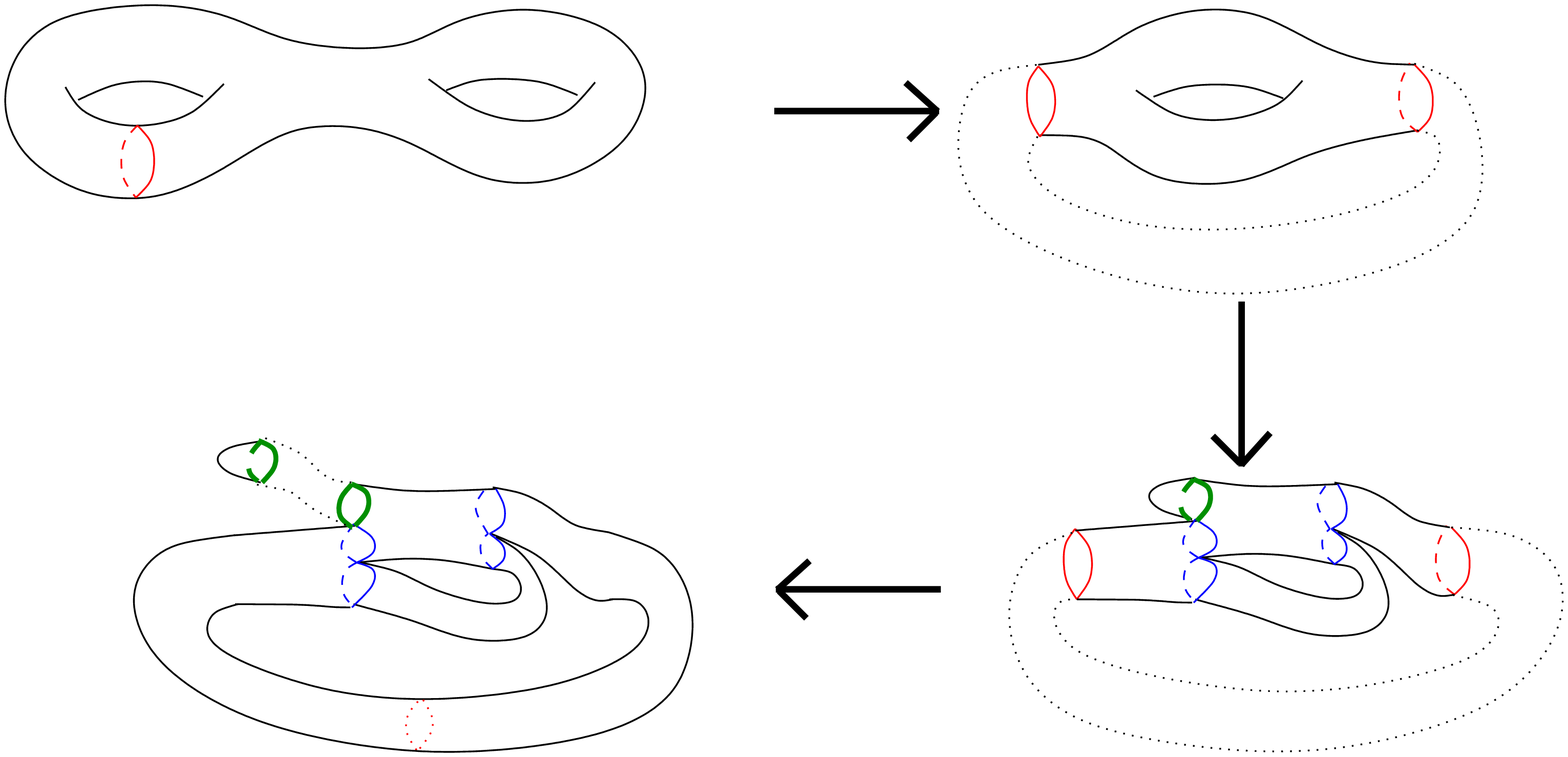}{\label{figDiscsF2} Example of a worldsheet in  ${\cal M}_{2,0}$. Chose an arbitrary point on the worldsheet, and cut the worldsheet along the horizontal trajectory going through that point. One may get a worldsheet in ${\cal M}_{1,2}$ or ${\cal M}_{1,1}\times{\cal M}_{1,1}$. Assume that we are in the ${\cal M}_{1,2}$ case. A worldsheet in ${\cal M}_{1,2}$, is obtained by gluing two propagators and 2 cylinders. The internal propagator has its initial boundary on a critical trajectory, and its boundary must be a circle, so that we need to close one half of it by a disk. This means that there is a disc on our initial worldsheet. Now, cut the initial worldsheet along the critical trajectory of the disc boundary. One gets a decomposition of our initial worldsheet into a disc and a worldsheet in ${\cal M}_{2,1}$.}

\section{Reconstruction of the spectral curve}

So far, we have defined generating functions $\widetilde W^{(g)}_k$ as formal series of complex formal variables $z_i$'s.
Here, we shall glue all patches in order to get generating functions globally defined on a Riemann surface $\widehat \curve$.
Then, afterwards, we shall show that this Riemann surface $\widehat \curve$, is actually the same as the underlying Riemann surface of our starting point spectral curve $\widehat \curve = \curve$.

\subsection{The disc amplitude}

Consider a Hurwitz space $(\widehat\curve, \hat x)$,  given by a Riemann surface $\widehat\curve$ of genus $\genus$ (the same genus as $\curve$ defining the spectral curve of the integrable system, i.e. the dimension of the Jacobian), and a projection $\hat x:\widehat\curve \to \mathbb CP^1$, with as many simple ramification points as the $a_i$'s.

Assume here that $d\hat x$ is a meromorphic form, whose zeroes are simple, and are labeled by the $a_i$'s.

The degree of $d\hat x$ (the number of poles with multiplicities) is then $\#\{a_i\} - (2\genus-2)$.

\medskip

Notice that if $\zeta\in \widehat\curve$ is a point on $\widehat\curve$ near $a_i$, then $z=\sqrt{\hat x(\zeta)-\hat x(a_i)}$ is a  local coordinate on $\widehat\curve$ near $a_i$, and we have:
\beq
\hat x(\zeta)=\hat x(a_i)+z^2.
\eeq
The two branches $\zeta$ and $\bar\zeta$ coming together at the ramification point, correspond to $z$ and $-z$ respectively , i.e.
\beq
\hat x(\bar \zeta)=\hat x(\zeta)
\virg
\sqrt{\hat x(\bar\zeta)-\hat x(a_i)}=-\,\sqrt{\hat x(\zeta)-\hat x(a_i)}.
\eeq

\medskip

Now, let us try to find an analytical function $\hat y$, defined on an open domain of $\widehat\curve$, containing all branchpoints, and such that in the vicinity of branchpoint $a_i$, we have in the local coordinate $z=\sqrt{\hat x-\hat x(a_i)}$:
\beq
2z\,(\hat y(\zeta)-\hat y(\bar\zeta))\,= \widetilde W^{(0)}_{1}\left(a_i,\sqrt{\hat x(\zeta)-\hat x(a_i)}\right)+\widetilde W^{(0)}_{1}\left(a_i,-\sqrt{\hat x(\zeta)-\hat x(a_i)}\right).
\eeq

\medskip

The existence of such a function globally defined of $\widehat\curve$ is not obvious, so, let us assume that we are in a situation where it does exist\footnote{As it is pointed out in the next section, in physics, one goes the other way round: from a problem in physics or mathematics, one derives such an integrable system by the computation of the
simplest observables whose generating function satisfy an equation defining the spectral curve. The existence of
such a globally defined function is thus ensured from the beginning.}.

Then, it is clear that if such a function does exist, it is not unique. Indeed, one may add to $\hat y$ any rational function of $\hat x$: locally near any branchpoints $a_i$ we can add any even function of $z=\sqrt{\hat x-\hat x(a_i)}$.
In other words, $\hat y$ is not unique, it is defined up to an additive rational function of $\hat x$.

So, let us assume that we have chosen one function $\hat y$ on $\widehat\curve$.

\subsection{The 2-point function}

The cylinder amplitude should satisfy the self-reproducing property, i.e. the fact that gluing a cylinder to a cylinder, is again a cylinder.
This should hold only if gluings are performed away from singularities (branchings or punctures).

In other words, for any $p_1,p_2,p_3$, and any $l_1$ and $l_2$:
\beq
\NN^{(0)}_2(p_1,l_1;p_2,l_2) = \int_0^\infty {dl\over l}\,\, \NN^{(0)}_2(p_1,l_1;p_3,l)\,\NN^{(0)}_2(p_3,l;p_2,l_2)
\eeq
(we divide by $l$ so that the marked point is not counted twice).

If we want to translate that into a global property for the generating function, we write:
\beq
\widetilde W^{(0)}_2(p_1,z_1;p_2,z_2) = \Res_{z\to z_1}\,\, \widetilde W^{(0)}_2(p_1,z_1;p_2,z)\,dz\,\, \int^z\widetilde  W^{(0)}_2(p_1,z';p_2,z_2)\,dz'.
\eeq
which implies that near $z_1=z_2$ we must have:
\beq
\widetilde W^{(0)}_2(p_1,z_1;p_2,z_2) \sim {1\over (z_1-z_2)^2}
\eeq
in any local variable $z$, and for any $p_1,p_2$ close enough so that we can choose the same local  variable for $z_1$ and $z_2$.

\medskip

On a Riemann surface $\widehat\curve$, there exists a Bergman kernel, i.e. a symmetric $2$-form $B(\zeta_1,\zeta_2)$, having a double pole at $\zeta_1=\zeta_2$, with vanishing residue, and no other pole, and normalized so that:
\beq
B(\zeta_1,\zeta_2) \mathop{{\sim}}_{z_1\to z_2}\,\, {dz(\zeta_1)\,dz(\zeta_2)\over (z(\zeta_1)-z(\zeta_2))^2} + {\rm regular}
\eeq
where $\zeta_i\in \widehat\curve$ stands for a point of $\widehat\curve$, and $z(\zeta_i)\in \mathbb C$ is any local coordinate on $\widehat\curve$.

Such a form is the definition of the "Bergman kernel", see \eq{eqdefBergman}.
One ambiguity remains. In order to be uniquely defined, the Bergman kernel is defined normalized on a chosen symplectic homology basis of non-contractible cycles on $\widehat\curve$, see \eq{eqBpluskappa}.
What we can say, is that given a curve $\widehat\curve$, and functions $\hat x$ and $\hat y$, there is not in general a unique choice of a Bergman kernel.
We need another data, independent from the disc amplitude, and this data is a $\genus\times \genus$ symmetric complex matrix $\kappa$, as in \eq{eqBpluskappa}.

\bigskip

Let us assume that we can choose a Bergman kernel, globally defined on $\widehat\curve$, such that near every two branchpoints $a_i, a_j$ we have (where $z_i(\zeta)=\sqrt{\hat x(\zeta_1)-\hat x(a_i)}\,$):
\beq
\widetilde W_2^{(0)}\left(a_i,z_i(\zeta_1);a_j,z_j(\zeta_2)\right)\,dz_i(\zeta_1)\,dz_j(\zeta_2) = B(\zeta_1,\zeta_2).
\eeq

\subsection{Higher topology amplitudes}

Then, the topological recursion implies by an easy recursion, that all generating functions of the type
\beq
\widetilde W_k^{(g)}(a_{i_1},z_1;a_{i_2},z_2;\dots;a_{i_k},z_k)
\eeq
can be defined globally as meromorphic differential forms on the curve $\widehat\curve$:
\bea
&& \widetilde W_k^{(g)}(a_{i_1},z_{i_1}(\zeta_1);a_{i_2},z_{i_2}(\zeta_2);\dots;a_{i_k},z_{i_k}(\zeta_k)) \,\, dz_{i_1}(\zeta_1)\dots dz_{i_k}(\zeta_k) \cr
&=& W_k^{(g)}(\zeta_1,\dots,\zeta_k).
\eea
Since at each step in the recursion, residues are computed at branchpoints, this implies that the forms $W_k^{(g)}$'s have poles only at branchpoints.

\medskip

Therefore, we have obtained that the generating functions "counting" worldsheets in ${\cal M}_k^{(g)}$, are the correlators of \cite{EOFg} obtained by the topological recursion from the spectral curve $(\widehat\curve,\hat x,\hat y)$.

\medskip

Similarly, generating functions $F_g$ of worldsheets with no boundary, are the symplectic invariants
$F_g$ of the spectral curve $(\widehat\curve,\hat x,\hat y)$.

\subsection{Reconstructing the integrable system}
\label{secreconstructspcurvedisc}

It was claimed in \cite{EOFg}, and made precise in \cite{EMnonpert} that, out of the symplectic invariants and correlators of a spectral curve $(\widehat\curve,\hat x,\hat y)$, it is possible to construct a formal Tau-function (formal function of $g_s$), satisfying Hirota's equations:
\beq
{\cal T}(g_s) = \exp{\left({\displaystyle \sum_g} g_s^{2g-2}\, F_g(\widehat\curve,\hat x,\hat y)\right)}\,\,\, \Big( 1+ g_s(\Theta'\,F_1' + {1\over 6}\, \Theta'''\, F_0''') +\dots \Big)
\eeq
(see the exact expression in \cite{EMnonpert}).

This Tau function defines an integrable system, whose classical limit at $g_s\to 0$, is a classical integrable system of spectral curve $(\widehat\curve,\hat x,\hat y)$.

\medskip

Therefore, starting from some integrable system, whose classical limit has a spectral curve $(\curve,x,y)$, we have, by using the flat connection, defined a "topological string theory", whose amplitudes define themselves another integrable system whose classical spectral curve is $(\hat\curve,\hat x,\hat y)$.
Those two integrable systems encode the same information, they are dual to one another, and therefore they should have the same spectral curve up to symplectic transformations.

In other words, up to symplectic transformations, the spectral curve $(\widehat\curve,x,y)$ is the spectral curve of the initial integrable system.
In particular the underlying Riemann surface is the same:
\beq
(\widehat\curve,\hat x,\hat y) \equiv (\curve,x,y)\quad {\rm modulo\, symplectic\, transformations}.
\eeq

\bigskip

Therefore, a posteriori, we determine the disc amplitudes by $W_1^{(0)} = ydx$, and the cylinder amplitudes as the Bergman kernel on $\curve$.

\subsection{Ambiguity of the construction}
\label{secambiguity}

Let us notice that we have made some arbitrary choices at some points.

\subsubsection*{Framing and choice of an integrable system}

 First, given a spectral curve, we have chosen a realization of a classical integrable system attached to it.
This choice is not unique. In particular, given a curve $\curve$, we have chosen a projection $x:\curve \to  \mathbb CP^1$, to obtain a Hurwitz space, and we have chosen a function $y$ (for instance we have seen that we have the freedom to add to $y$ any rational function of $x$).
The Tau-function (and the $F_g$'s) are unchanged if we choose another representant of the same integrable system, in particular we may change $(x,y)\to (\td x,\td y)$ such that $d\td x\wedge d\td y=dx\wedge dy$, without changing the $F_g$'s and the Tau-function.
But when doing that, we change the open amplitudes $W_n^{(g)}$'s.

\smallskip
This ambiguity can be thought of as a choice of "framing".

\smallskip

For example, in the context of topological B strings, the mirror spectral curve is of the form
\beq
H(\ee{x},\ee{y})=0
\eeq
where $H$ is a polynomial.
For any integer $f$, changing $y\to y+f\,x$ doesn't change the $F_g$'s and $W_n^{(g)}$'s, and it corresponds to
\beq
\ee{y} \to \ee{y}\,\,(\ee{x})^f
\eeq
which is a well known framing transformation in topological B strings.

\subsubsection*{Polarization and modularity}

 Then, we have chosen a polarization vector $\vec v$ in the lattice $\mathbb Z^\genus+\tau\mathbb Z^\genus$. This polarization vector can be viewed as a characteristics in the Jacobian. It can be linked to a choice of a symplectic basis of non-contractible cycles ${\cal A}_i\cap {\cal B}_j=\delta_{i,j}$ on the curve $\curve$. Indeed, a modular transformations (i.e. a change of symplectic basis $({\cal A}_i,{\cal B}_j)$), is equivalent to a $Sl_\genus(\mathbb Z)$ transformation of the lattice, and can change $\vec v$ to any other lattice vector.
Notice that we have the same ambiguity in the choice of a Bergman kernel on $\curve$. In other words, the choice of the Bergman kernel, should be linked to the polarization vector $\vec v$.

\smallskip
The open amplitudes $W_n^{(g)}$'s and closed amplitudes $F_g$'s, depend explicitly on the choice of polarization, i,e, they are not invariant under modular transformations.
It was proved in \cite{EOFg,eynhaeq}, that modular transformations of the $F_g$'s and $W_n^{(g)}$'s, obey the formalism of BCOV \cite{BCOV}, and are given by the diagrammatic rules of \cite{AganGIC}.

BCOV \cite{BCOV}, and \cite{AganGIC} noticed that those modular changes of the $F_g$'s can be canceled by adding some non-holomorphic terms to them.
It was long debated what the role of those non-holomorphic terms could be, and in particular, there is no such non-holomorphic terms in the Chern-Simons field theory which is supposed to be dual to the topological string B-model.

Recently it was discovered \cite{EMnonpert} that the modular changes of the $F_g$'s can also be canceled by some holomorphic non-perturbative terms. In other words, the holomorphic anomaly is merely an artifact of perturbative expansion.

In fact those non-perturbative terms are essential to make the whole partition function satisfy Hirota equations and be the Tau-function of an integrable system, and we used them in section \ref{secreconstructspcurvedisc} above.

\subsubsection*{Non-perturbative part}

The whole string theory partition function thus contains a perturbative part, given by the $F_g$'s, which count open strings of finite genus $g$ with a given polarization, and a non-perturbative part, which restores modular invariance, background independence and integrability.
It would be interesting to understand what worldsheets  the non-perturbative terms count.
A guess is that they count worldsheets which are non-compact Riemann surfaces, of infinite genus.

\section{Other  expansions, Lagrange inversion}
\label{secexpansions}

In the previous section, we have constructed some string amplitudes $W_n^{(g)}$ which are meromorphic forms intrinsically defined on the curve $\curve$. They were constructed from a spectral curve $\spcurve=(\curve,x,y)$.

\medskip

By construction, when we expand $W_n^{(g)}(\zeta_1,\dots,\zeta_n)$ as Laurent series in terms of the local variables $z_i=\sqrt{x(\zeta_i)-x(a)}$ near a branchpoint $a$, the coefficients of the expansion of $W_n^{(g)}$ are Laplace transforms of the generating functions counting worldsheets of topology $(g,n)$ and with boundaries of given lengths:
$$
W_n^{(g)}(\zeta_1,\dots,\zeta_n) = \sum_{k_1,\dots,k_n} A_{a,k_1;\dots;a,k_n}^{(g)}\,\, \prod_i (k_i+1) z_i^{-k_i-2}\, dz_i
$$
where
\beq
\NN_n^{(g)}(a,l_1;\dots;a,l_n) = \sum_{k_1,\dots,k_n} A_{a,k_1;\dots;a,k_n}^{(g)}\,\, \prod {l_i^{k_i+1}\over k_i !}.
\eeq
Notice that, when $2-2g-n<0$, from the general property of the topological recursion we have that  $A_{a,k_1;\dots;a,k_n}^{(g)}=0$ if $k_i>6g-6+2n$, but the $k_i$'s can take negative values down to $-\infty$.
Notice also, that if $k_i>0$ then $k_i$ must be even, and for $k_i<0$, there is no parity restriction.

\bigskip

For $2-2g-n<0$, the topological recursion ensures that the $W_n^{(g)}$ are meromorphic forms with even poles only at branchpoints, and of degree at most $6g-4+n$, we may decompose them on a basis of such meromorphic forms.
Consider:
$$
B_{a_i,n}(\zeta) = \Res_{\zeta'\to a_i} \, B(\zeta,\zeta')\,\, (x(\zeta')-x(a_i))^{-n-{1\over 2}}
$$
which is a meromorphic form, whose only pole is at $a_i$, and which behaves like:
$$
B_{a_i,n}(\zeta) \sim (2n-1)\,\,{dz\over z^{2n+2}} + {\rm reg}
$$
where we recall that $z=\sqrt{x(\zeta)-x(a_i)}$.

\medskip
Therefore we can write $W_n^{(g)}$ as a finite linear combination of $B_{a_i,k}$'s:
\beq
W_n^{(g)}(\zeta_1,\dots,\zeta_n) = \sum_{i_1,k_1;\dots;i_n,k_n} A_{a_{i_1},2k_1;\dots;a_{i_n},2k_n}^{(g)}\,\, \prod_{j=1}^n B_{a_{i_j},k_j}(\zeta_j),
\eeq
which is a finite sum since each $k_j$ is between:
\beq
0\leq k_j\leq 3g-3+n.
\eeq

\subsection{Expansion near other points}

One may also choose to expand $W_n^{(g)}(\zeta_1,\dots,\zeta_n)$ as a Taylor or Laurent series near any other point $p$ (not necessarily a branchpoint), and in powers of any other variable $\td z=f(\zeta)$:
$$
W_n^{(g)}(\zeta_1,\dots,\zeta_n) = \sum_{k_1,\dots,k_n} {\td A}_{p,k_1;\dots;p,k_n}^{(g)}\,\, \prod \td z_i^{-k_i-1}\, d\td z_i .
$$

In general, the coefficients ${A}_{k_1,\dots,k_n}^{(g)}$ of Laurent expansion of a function in terms of one local variable $z_i$, are related to the coefficients ${\td A}_{k_1,\dots,k_n}^{(g)}$ of Laurent expansion  in terms of another variable $\td z_i$, by the Lagrange inversion formula, which just amounts to compute the residues:
\bea
{\td A}_{p,k_1;\dots;p,k_n}^{(g)}
&=& \Res_{\zeta_1\to p}\dots \Res_{\zeta_n\to p}\,\, W_n^{(g)}(\zeta_1,\dots,\zeta_n)\,\, \prod_{i=1}^n\, f(\zeta_i)^{k_i}.
\eea

Here, it suffices to compute the Taylor or Laurent series expansion of the basis forms $B_{a_i,n}$ in the parameter $\td z$:
$$
B_{a_i,n}(\zeta) = \sum_k B_{a_i,n;p,k}\,\, \td z^{-k-1}\,d\td z
\qquad , \qquad
B_{a_i,n;p,k} = \Res_{\zeta\to p}\,\, B_{a_i,n}(\zeta)\,\, f(\zeta)^k.
$$

And that gives:
\beq\label{ELSVgeneralized}
{\td A}_{p_1,k_1;\dots;p_n,k_n}^{(g)}
= \sum_{i_1,m_1;\dots;i_n,m_n} A_{a_{i_1},2m_1;\dots;a_{i_n},2m_n}^{(g)} \prod_{j=1}^n\, B_{a_{i_j},m_j;p_j,k_j}
\eeq

\subsubsection*{Remark on Cut and Join equations}

The topological recursion implies some recursive equations among the coefficients $A^{(g)}_{a_{i_1},k_1;\dots;a_{i_n},k_n}$, and therefore, through equation \eq{ELSVgeneralized}, they imply some relationships among the coefficients ${\td A}_{p_1,k_1;\dots;p_n,k_n}^{(g)}$.

Those relationships can be thought of as "{\bf Cut and Join}" equations.

Indeed, we shall see below, that for rather canonical choices of $f(\zeta)$, they correspond to Tutte's equations for discrete surfaces, to Cut and Join equations for Hurwitz numbers, or to Mirzakhani or Virasoro equations for intersection numbers of tautological classes.

\medskip
In fact, in most of the known applications of the topological recursion in physics and mathematics,
some recursion relations (Cut and Join, or Tutte) are known in terms of the coefficients ${\td A}_{p_1,k_1}^{(0)}$ and not directly $A^{(0)}_{a_{i_1},k_1}$, i.e. in terms of the Laurent expansion of a local coordinate of the spectral curve typically near a pole or a logarithmic singularity (see the
examples in  the next section), not near branchpoints.

\subsection{Some canonical choices of expansions}

\subsubsection{Meromorphic case}

It was observed in \cite{EOFg}, that, if the form $ydx$ is meromorphic, the coefficients of its Laurent series expansion near its poles, are the KP times:
\beq
y(\zeta)dx(\zeta) \mathop{{\sim}}_{\zeta\to p_j}  \sum_{k=0}^{d_j}\,\,  t_{j,k}\,\, z_{p_j}(\zeta)^{-k-1}\, d_{p_j}(\zeta)
\eeq
where $d_i$ is the degree of the pole of $ydx$ at $p_j$, and $z_{p_j}(\zeta)$ is a local coordinate near $p_j$ given by $z_{p_j}(\zeta)=x(\zeta)^{-1/\deg_{p_j} x}$ if $x$ has a pole at $p_j$, and $z_{p_j}(\zeta)=x(\zeta)-x(p_j)$ if $x$ has no pole at $p_j$.

This shows that a natural choice of expansion is to choose $\td z=f(\zeta)=1/(z_{p_j}(\zeta))$ near a pole $p_j$.
If we apply the Lagrange inversion formula as above, and expand the $W_n^{(g)}$'s in Laurent series in the variables $\td z$, we shall get a natural expansion in terms of KP times.

\smallskip
This kind of expansion is deeply related to the Frobenius manifold structure of moduli spaces of worldsheets.

\subsubsection*{Example: maps, discrete surfaces, 1-matrix model}

The formal 1-matrix model provides generating functions for counting maps (maps in the sense of combinatorics, i.e. graphs embedded on Riemann surfaces, also called discrete surfaces or ribbon graphs).
For the 1-matrix model, the spectral curve is algebraic, and $x(\zeta)$ is a meromorphic function, with two simple poles.
Moreover, the spectral curve is hyperelliptical, there is an involution $\zeta\to \bar\zeta$ on $\curve$, for which $x(\bar \zeta)=x(\zeta)$, and it turns out that all stable $W_n^{(g)}$'s are odd under that involution.
This implies that computing the Laurent expansion at one pole of $x$ is equivalent (up to a sign $(-1)^n$) to computing it at the other pole.

In that case we choose $f(\zeta)=x(\zeta)$, and we write:
\beq
W_n^{(g)}(\zeta_1,\dots,\zeta_n) = \sum_{k_1,\dots, k_n} {\td A}^{(g)}_{k_1,\dots,k_n}\,\, \prod_i x_i^{-k_i-1}\,dx_i.
\eeq
It is well known that the coefficients ${\td A}^{(g)}_{k_1,\dots,k_n}$ are the generating functions which count discrete surfaces of genus $g$, with $n$ marked faces of lengths $k_1,\dots,k_n$ and a marked edge on each marked face:
$$
{\td A}^{(g)}_{k_1,\dots,k_n} = \sum_{m} {1\over \#{\rm Aut}(m)}\,\, \prod_j t_j^{n_j(m)}
$$
where the sum is over the set of maps (or discrete surfaces) $m$ of genus $g$ and $n$ marked faces of lengths $k_1,\dots,k_n$ and a marked edge on each marked face.
$n_j(m)$ denotes the number of unmarked faces of $m$ of valence $j$, and $t_j$ is the KP-time, found from the Laurent series expansion of $ydx$ at large $x$:
$$
y\,dx = \sum_j t_j\, x^{j-1}\,dx.
$$

It is also known that the topological recursion, written in terms of ${\td A}^{(g)}_{k_1,\dots,k_n}$, reduce to Tutte's equations \cite{tutte1,tutte2} for discrete surfaces (also called loop equation under their matrix model's representation \cite{Migdalloop}).
In particular, let us derive them for the simplest case of the disc amplitudes,
$n=1$ and $g=0$.

Tutte's recursion consists in removing the marked edge from the marked face of degree $k+1$ and enumerating all possible maps obtained in this way:
\begin{itemize}
\item either the other side of the marked edge is in another face of
degree $l$ and removing it gives a map with one marked face of degree $k+l-1$;

$$
{\mbox{\epsfxsize=8.truecm\epsfbox{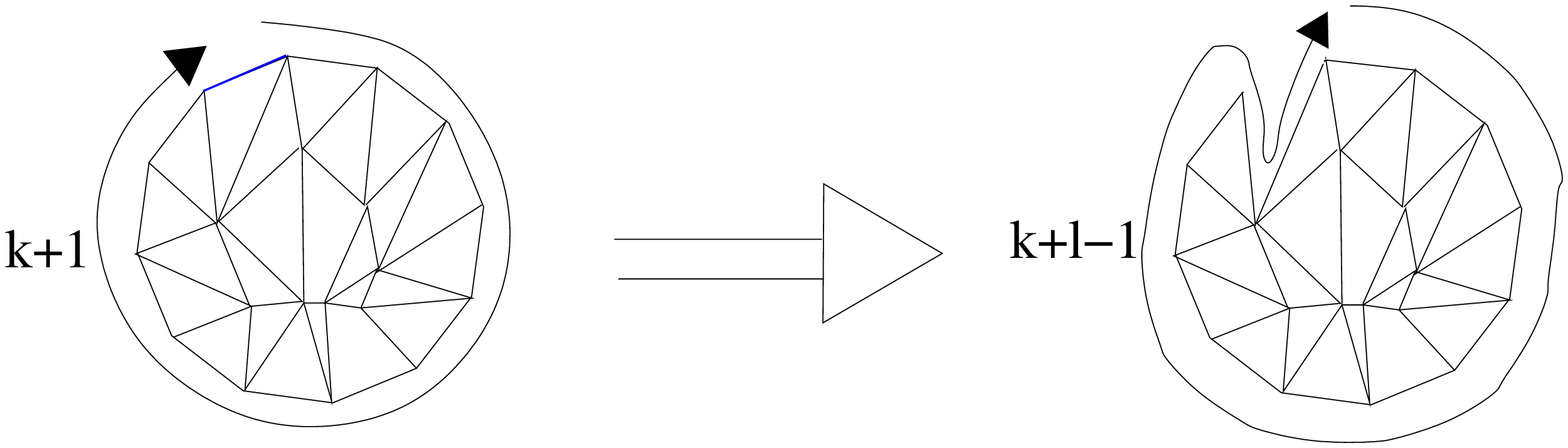}}}
$$

\item either the other side of the marked edge is the same face and removing this edge disconnects the map into two
components: one with one marked face of degree $l\leq k$ and one with one marked face of degree $k-l-1$.

$$
{\mbox{\epsfxsize=10.truecm\epsfbox{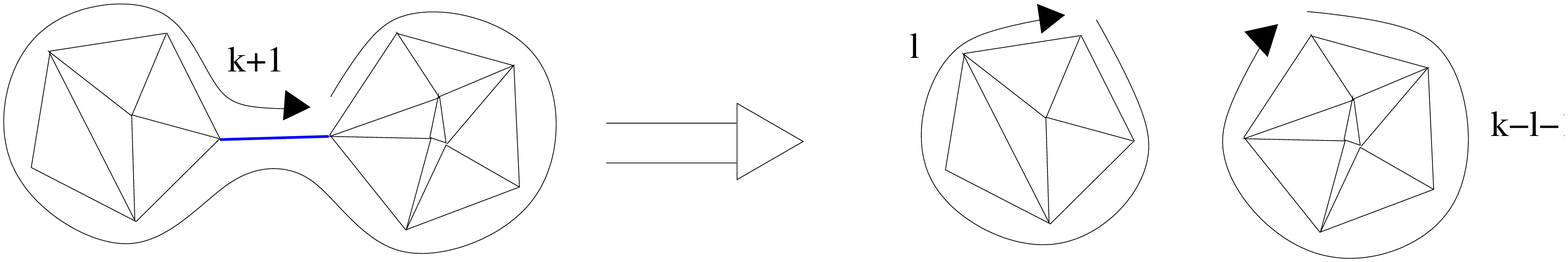}}}
$$

\end{itemize}
This procedure is bijective and gives the following relation:
$$
-t_2\,{\td A}^{(0)}_{k+1} = \sum_{l=0}^{k-1} {\td A}^{(0)}_{l}{\td A}^{(0)}_{k-l-1} + \sum_{l\neq 2} t_l {\td A}^{(0)}_{k+l-1}
$$
which can indeed be thought of as a cut-and-join relation for the discs.
In terms of the generating function $W_1^{(0)}(x) = {t\,dx\over x}+\sum_{k\geq 1} {\td A}^{(0)}_{k} x^{-k-1}\,dx = y dx$, it reads
$$
y^2 + y V'(x) = P(x) = \left[y\,\, V'(x)\right]_+
$$
where $V'(x) = \sum_j t_j x^{j-1}$, and $P(x)=\left[f(x)\right]_+$ is the positive part of the Laurent expansion in $x$ of $f(x)$, thus $P(x)$ is a polynomial of $x$.

This implies that the disc amplitude satisfies an algebraic equation, and we have
$$
y= {1\over 2}\, \left( V'(x) \pm \sqrt{V'(x)^2-4 P(x)}\right).
$$
This equation defines the spectral curve of
this matrix model, or enumerative problem of maps.

Using this procedure to remove one edge of maps of arbitrary topology gives more general Tutte's
equations which define recursively the coefficients ${\td A}^{(g)}_{k_1,\dots,k_n}$: it was proved in \cite{eynloop1mat} these recursions are equivalent to the topological recursion for the $W_n^{(g)}$'s, and thus that the ${\td A}^{(g)}_{k_1,\dots,k_n}$ are the result
of the Lagrange inversion formula on the topological recursion for the coefficients ${A}^{(g)}_{k_1,\dots,k_n}$.

\subsection{Case of non-meromorphic singularities}
\label{secexpansionlog}

If $ydx$ has non meromorphic singularities, then $f(x)=$power of $x$ can't be a good expansion parameter.

For example, assume that $x$ has a logarithmic singularity, i.e. assume that $\ee{x}$ has a meromorphic singularity at some pole $p$.
Then, it is natural to expand in powers of $\ee{x}$.

This type of logarithmic singularity occurs for applications to topological strings, because the spectral curve is of the form $H(\ee{x},\ee{y})=0$ where $H$ is a polynomial. In that case, $\ee{x}$ and $\ee{y}$ are meromorphic functions on $\curve$.

\medskip
Then, one can compute, through formula \eq{ELSVgeneralized}, the coefficients of the expansion
\beq
W_n^{(g)}(\zeta_1,\dots,\zeta_n) = \sum_{k_1,\dots, k_n} {\td A}^{(g)}_{k_1,\dots,k_n}\,\, \prod_i \ee{k_i x_i}\,dx_i,
\eeq
in terms of the $A^{(g)}_{a_{i_1},k_1;\dots;a_{i_n},k_n}$ computed at branchpoints.

The relationship between the coefficients ${\td A}^{(g)}_{k_1,\dots,k_n}$ which compute numbers of worldsheets having given perimeters near the log singularities of $x$, and the coefficients $A^{(g)}_{a_{i_1},k_1;\dots;a_{i_n},k_n}$ computing worldsheets having given perimeters near branchpoints, can be thought of as a kind of generalization of ELSV formula \cite{ELSV}.

\medskip

Also in that case, the topological recursion written for the coefficients ${\td A}^{(g)}_{k_1,\dots,k_n}$, can be viewed as a generalization of the cut and join equations \cite{cutandjoin1,cutandjoin2}.

\subsubsection*{Example: Hurwitz numbers and ELSV}

The spectral curve for Hurwitz numbers is related to the Lambert function $y=L(\ee{x})$:
$$
\ee{x} = y\ee{-y}.
$$
This means that $y$ is a meromorphic function on $\mathbb CP^1$, with a pole which we choose to be at $0$:
$$
y(\zeta) = 1-{1\over \zeta}
$$
and $x$ is not meromorphic, it has a pole at $z=0$ and a log singularity at $\zeta=1$ and $\zeta=\infty$:
$$
x(\zeta) = -1+{1\over \zeta}+\ln{(1-{1\over \zeta})} = -1-\sum_{k=2}^\infty {1\over k\, \zeta^k}.
$$

There is a unique branchpoint at $\zeta=a=\infty$, where $dx$ vanishes.
The local parameter near the branchpoint is:
$$
z=\sqrt{-x-1} = {1\over \zeta\,\sqrt 2}\,\left( 1+{1\over 3\zeta}+ O(\zeta^{-2}) \right).
$$

For any spectral curve with only one branchpoint, it is easy to write the expansion of $W_n^{(g)}$ near the branchpoint $\zeta\to\infty$, i.e. $z\to 0$, in terms of intersection numbers of tautological classes, see \cite{BEMS}. Indeed, since the topological recursion computes residues only at the branchpoint, we need only to know the Taylor expansion of $y(\zeta)-y(\bar \zeta)$ near the branchpoint:
\beq
y(\zeta)-y(\bar\zeta) = \sum_k t_{2k+1} (x(\zeta)-x(\infty))^{k-{1\over 2}}
\eeq
and this quantity is exactly the Kontsevich spectral's curve with times $t_k$.
The $W_n^{(g)}$'s can then be expressed in terms of intersection numbers of $\psi$ and $\kappa$ classes, see \cite{eynMgnkappa}.

\medskip

On the other side, it is known that the expansion near $\zeta=1$ in terms of $\ee{x}$ gives the Hurwitz numbers \cite{BouchardMarino}.

In that case, the Lagrange inversion formula can be viewed as the ELSV formula \cite{ELSV},
and the topological recursion in terms of Hurwitz numbers can be viewed as the cut and join equations of Goulden-Jackson-Vakil \cite{cutandjoin1,cutandjoin2}.

\medskip
Let us point out that the spectral curve was obtained by these cut-and-join equations for the disc amplitudes, as in
the discrete surfaces case. Hurwitz numbers $h_{g,\mu}$ enumerate coverings of $\mathbb{CP}^1$ by genus $g$
surfaces ramified over infinity with profile $\mu$ and at most simply ramified anywhere else. The Riemann
Hurwitz formula fixes the number of simple ramification points away from infinity, to be $2g-2+l(\mu)+|\mu|$. One can get a recursion on the Hurwitz
numbers $h_{g,\mu}$ by removing (or resolving) one simple ramification point from the such coverings and enumerating all possible
changes of the ramification profile $\mu$ compatible with such a resolution. In particular, for $g=0$
and $\mu = \left\{n\right\}$, this procedure gives cut-and-join equations defining recursively the "disc amplitudes"
$h_{0,n}$. Let us define $H_{g,\mu}:={\left|\hbox{Aut}(\mu)\right| \over (|\mu| + l(\mu)+2g-2)} h_{g,\mu}$,
the cut-and-join equation reads:
\beq\label{recW1}
{(n-1) \over n} {H}_{0,n} = {1 \over 2} \sum_{k=1}^{n-1} {{H}_{0,k} } {{H}_{0,n-k}}.
\eeq
This can be turned into a differential equation for the generating function $y=W_1^{(0)}(x)/dx = \sum_n {H}_{0,n} e^{nx}$:
\beq
2 y - y^2  = \int {y \over x} d\left(e^x\right)
\eeq
which is a mere rewriting of the Lambert equation $y = e^{x+y}$.
The cut-and-join equations thus allow to find the spectral curve associated to this enumerative problem
in the patch near $\zeta=1$.
The other cut-and-join equations are then obtained by the Lagrange inversion formula
from $\zeta=1$ to $\zeta=\infty$ and lead to the topological recursion.

\subsection{Decomposition on branchpoints}

When the spectral curve has a single branchpoint $a$, since the topological recursion computes residues only at the branchpoint, we need only to know the Taylor expansion of $y(\zeta)-y(\bar \zeta)$ near the branchpoint ($z=\sqrt{x(\zeta)-x(a)}$):
$$
y(\zeta)-y(\bar\zeta) = 2z-\sum_{k\geq 1} t_{2k+1}\,\, z^{2k-1}
$$
and we recognize exactly the Kontsevich spectral's curve with times $t_k={1\over N}\Tr \Lambda^{-k}$ (see \cite{EOFg,eynMgnkappa}).
The $W_n^{(g)}$'s and $F_g=W_0^{(g)}$'s of the Kontsevich integral can then be expressed in terms of intersection numbers of $\psi$ and $\kappa$ classes, see \cite{eynMgnkappa}, and the result is:
 \bea
  W_{n}^{(g)}(\zeta_1,\dots,\zeta_n)
&=& 2^{- d_{g,n}}(2\ee{-\td{t}_0})^{2-2g-n}\!\!\!\!  \sum_{d_0+d_1+\dots+d_n=d_{g,n}}\,
 \sum_{k=0}^{d_0} {1\over k!}\,\sum_{b_1+\dots+b_k =d_0, b_i>0}  \cr
 && \qquad \qquad \prod_{i=1}^n {2d_i+1!\over d_i!}\, {dz_i\over z_i^{2d_i+2}}\,\, \prod_{l=1}^k \td{t}_{b_l} <\prod_{l=1}^k \kappa_{b_l} \prod_{i=1}^n  \psi_i^{d_i}>_{\ovl{\cal M}_{g,n}}  \cr
\eea
where $z_i=\sqrt{x(\zeta_i)-x(a)}$, and the dual times $\td t_k$ are related to the $t_k$'s by the following transformation:
$$
t_1=0
\virg
2-t_3=2\ee{-\td{t}_0}
$$
$$
f(z)  = \sum_{a=1}^\infty {(2 a+1)!\over a!}\,\,{t_{2a+3}\over 2-t_3} \,\, z^a
\quad \to \quad
\td{f}(z)= -\ln{(1-f(z))} = \sum_{b=1}^\infty \td{t}_b \,\, z^b  .
$$

For a given branchpoint $a$, we write:
$$
Z_{\rm Kontsevich}(a) = \ee{\sum_g F_g}.
$$

\medskip

When there are several branchpoints, it was shown in \cite{OrGiv}, following \cite{AMM1,AMM2} that the symplectic invariants $F_g$ can be obtained in terms of a product of Kontsevich integrals at each vertex:
\beq\label{eqgivental}
\ee{\sum_g F_g} = \ee{U_{\rm mixing}}\,\, . \,\prod_{a_i}\,\, Z_{\rm Kontsevich}(a_i),
\eeq
where $U_{\rm mixing}$ is a mixing operator:
$$
U_{\rm mixing} := {\displaystyle \sum_{i,j} } \oint_{a_i} \oint_{a_j} \hat{B}^{(i,j)}(\zeta_1,\zeta_2) \hat{\Omega}_i(\zeta_1)\hat{\Omega}_j(\zeta_2).
$$
with
$$
\hat{B}^{(i,j)}(\zeta_1,\zeta_2):= B(\zeta_1,\zeta_2) - {dz_i(\zeta_1) dz_j(\zeta_2) \over (z_i(\zeta_1)-z_j(\zeta_2))^2}
$$
and $\hat{\Omega}_i(\zeta_1)$ the differential operator:
$$
\hat{\Omega}_i(\zeta) := \sum_{k \geq 1} t_{k,i} z_i^k(\zeta) dz_i(\zeta) - {dz_i(\zeta) \over k z_i(\zeta)^k} {\partial \over \partial t_{k,i}},
$$
using the notations:
$$
z_i(\zeta):= \sqrt{x(\zeta)-x(a_i)}
\qquad \hbox{and} \qquad y(\zeta) - y(\ovl{\zeta}) = 2 z_i(\zeta) - \sum_{k\geq 1} t_{2k+1,i} z_i^{2k-1}(\zeta).
$$
The times $t_{k,i}$ are thus the Kontsevich  times of the $i$'th factor of the expression \eq{eqgivental}.

This mixing formula shows that all coefficients
$A^{(g)}_{a_{i_1},k_1;\dots;a_{i_n},k_n}$ can be written in terms of intersection numbers of tautological classes.

\bigskip

Therefore, the Lagrange inversion formula allows to express the coefficients ${\td A}^{(g)}_{a_{i_1},k_1;\dots;a_{i_n},k_n}$ corresponding to another expansion, in terms of intersection numbers.
This can be viewed as a generalization of the ELSV formula.

\section{Application: topological strings and BKMP conjecture}

The previous sections were aimed at general spectral curves. Here, we focus on spectral curves related to the mirror geometry of toric Calabi-Yau 3 folds, and the application to topological strings.

\smallskip

In \cite{BKMP}, Bouchard, Klemm, Mari\~no and Pasquetti conjectured that the open and closed amplitudes of type A
topological string theories on some Toric Calabi-Yau 3-folds coincide with the symplectic invariants
of their mirror B-model's target space. In this section, we explain why the computation of the Gromov-Witten
invariants (closed and open -- or relative as defined by \cite{mathematicalvertex}) reduces to the enumerative
problem solved earlier in this paper. Once again, this is not a proof but just a heuristic explanation of this
conjecture.

\subsection{Toric Calabi-Yau 3-folds and localization}

Let us consider a A-model topological string theory whose target space $\CYX$ is a toric Calabi-Yau three-fold.

One of the main features of this type of geometry is that it can be realized as the gluing of a set of
$\mathbb{C}^3$ patches, as a $\mathbb{T}^3$ fibration over a non-compact convex subspace of $\mathbb{R}^3$. Its
 geometry can be described by a Toric graph together with
the K\"ahler parameters of $\CYX$. The Toric graph corresponds to the degeneration of the $S^1$ fibers (of the
$\mathbb{T}^3$ fibration): the lines
of the Toric graph represent the locus where two out of the three $S^1$'s shrink to zero (see fig.\ref{figtoricgraph} for the simplest example  $\CYX=\mathbb{C}^3$).
In particular, the external legs of the Toric graph correspond to such degeneration along a non-compact direction:
such external legs are thus $\mathbb{C}$'s which can be compactified to $\mathbb{CP}^1$'s by including the point at
infinity.

This target space can be equipped with many possible Torus actions. Each of them leads to a different localization computation and thus, a different hamiltonian system.
In order to match the classical computations of topological string theories, let us choose one particular $S^1$-action on $\CYX$. For this purpose,
choose one external leg $L_{23}$ of the Toric graph and denote by $X_2$ and $X_3$ the two coordinate of $\CYX$
which vanish along this edge (they correspond to the two shrinking $S^1$'s along this external leg). Denote by $X_1$
the remaining coordinate of $\CYX$ which vanishes along the two other legs $L_{12}$ and $L_{13}$ of the vertex at which $L_{23}$ ends.

Now consider the Torus action on $\mathbb{C}^3$:
$$
T_{\rho_1,\rho_2,\rho_3}\, : \, (X_1,X_2,X_3) \rightarrow (t^{\rho_1} X_1 , t^{\rho_2} X_2 ,t^{\rho_3} X_3 ).
$$
One of the fixed points of $\CYX$ under this Torus action is the tip of $L_{23}$ defined by
$p_1:=(X_1\to \infty, X_2=0 , X_3 = 0)$. If $\rho_1=f$, $\rho_2=-f$ and $\rho_3=0$ for some
  integer $f$, this action reduces to a $S^1$-action which
we shall consider from now on.


\figureframex{15}{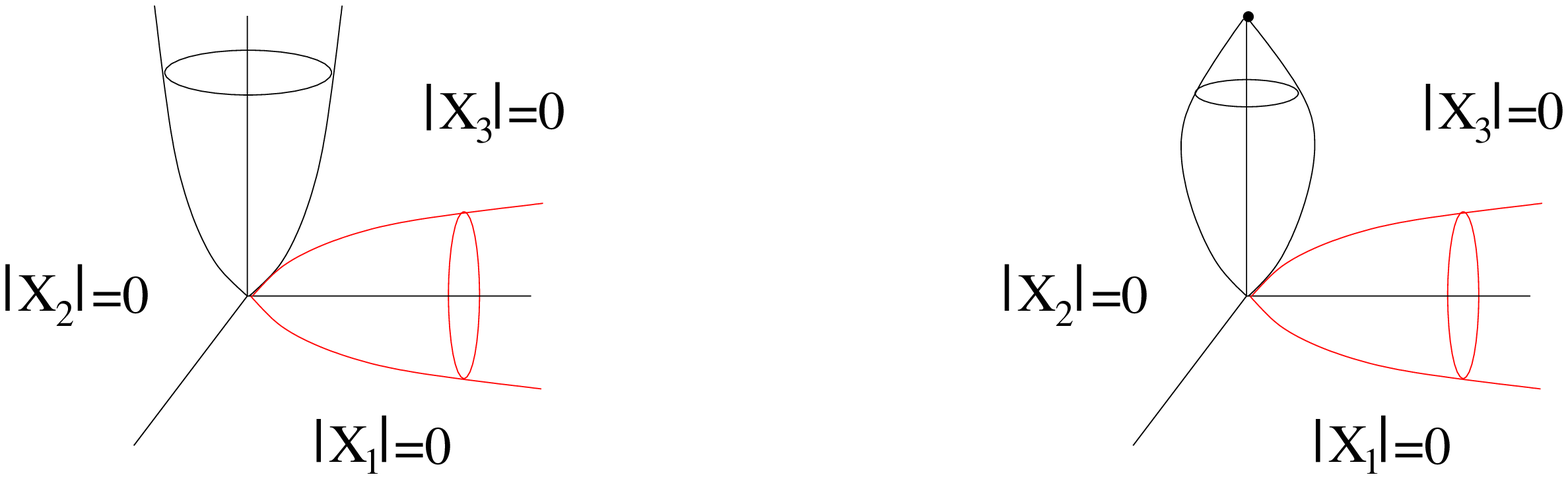}{\label{figtoricgraph} To the left: Toric graph of $\mathbb{C}^3$. The three planes $X_i=0$ intersect along the edges of the toric graph. The
  black circle represents a level of the Torus action with $\rho_1=f$, $\rho_2=-f$ and $\rho_3=0$ for some
  integer $f$:
  $(\left|X_1\right|=r,X_2=0,X_3=0)$ for some constant $r$. To the right: the same graph with the tip $p_1$ of the
external leg included to compactify the corresponding direction.}

Let us now enumerate holomorphic maps:
$$
m_{g,n}: \Sigma_g \to \CYX
$$
from a genus $g$ Riemann surface ${\Sigma_g}$ with $n$ marked points to $\CYX$, i.e. compute integrals of the form
$$
A_{n}^{(g)} = \int_{\overline{\cal M}_{g,n}} 1
$$
where $\overline{\cal M}_{g,n}$ is the compactification of the moduli space of such maps (see \cite{kontsevich} for
example).

The $S^1$-action on $\CYX$ induces a hamiltonian action on the moduli space $\overline{\cal M}_{g,n}$.
The Atiyah-Bott localization formula states that such integrals reduce to integrals over the fixed locus of the
Torus action:
\beq\label{eqlocalization}
A_{n}^{(g)} = \sum_{Fix_{g,n}} \int_{m_{g,n} \in Fix_{g,n}} {i^* m_{g,n} \over {\bf e}(N_{Fix_{g,n}/\overline{\cal M}_{g,n}})}
\eeq
where $\{Fix_{g,n}\}$ runs over the fixed locus of the Torus action and ${\bf e}(N_{Fix_{g,n}/\overline{\cal M}_{g,n}})$ is the Euler class of the normal bundle of $Fix_{g,n}$ in $\overline{\cal M}_{g,n}$\footnote{A rigorous
approach requires the introduction of an obstruction theory and consider the integrals over the virtual fundamental
class built out of it. Nevertheless, the fixed locus is the same as for the compactified moduli space: only the
integrant is changed in the localization formula (see \cite{bookmirror}).}.

This means that the observables reduce to a sum over all maps whose images in $\CYX$ are invariant under the
hamiltonian $S^1$-action counted with a weight  ${\bf e}(N_{Fix_{g,n}/\overline{\cal M}_{g,n}})$\footnote{This localization
formula is the sum over instantons of sec.\ref{secintroinstanton}. One instanton is a map stable under the Torus action, i.e. a fixed point of the moduli space.}. In particular, the marked points have to be mapped to fixed points of the $S^1$-action in $\CYX$ by maps in $Fix_{g,n}$.

For a fixed point $m_{g,n}\in Fix_{g,n}$, $m_{g,n}^{-1}$ allows
to lift the orbits of the Torus action in $\CYX$ to closed paths generating $\Sigma_g$: the level lines of
the $S^1$-action equip the worldsheet $\Sigma_g$ with a complex structure.
The coordinates in this foliation of the worldsheet are nothing but the action-angle variables of the Hamiltonian system under study, the $S^1$ circles are the horizontal trajectories of this hamiltonian action, at least locally near the point $p_1:=(X_1=\infty,X_2=0,X_3=0)$.

One has thus reduced the computation of $A_{n}^{(g)}$'s to the study of embeddings of worldsheets
into a Jacobian thanks to action-angle variables. This is exactly the problem studied in the previous section.

One can refine the description of the fixed locus of the $S^1$ action in $\overline{\cal M}_{g,n}$.
For such a Torus action, one can decompose $A_{n}^{(g)}$ by fixing the image of the marked points as well
as the local behavior of the enumerated maps. Let $p_1,\dots,p_m$ be fixed points of the $S^1$-action in $\CYX$. For a set of partitions $\mu_1,\dots,\mu_m$, let  $Fix_{g,\mu_1,\mu_2,\dots,\mu_m}$ be the set of maps $m_{g,\mu_1,\mu_2,\dots,\mu_m}$ in  $Fix_{g,\displaystyle \sum_{i=1}^m l(\mu_i)}$ such that $l(\mu_i)$ marked points
are mapped to $p_i$ with ramification profile $\mu_i$.
We finally denote
\beq
A_{\mu_1,\mu_2,\dots,\mu_m}^{(g)} = \sum_{Fix_{g,\mu_1,\mu_2,\dots,\mu_m}} \int_{m_{g,\vec{\mu}} \in Fix_{g,\mu_1,\mu_2,\dots,\mu_m}} {i^* m_{g,\vec{\mu}} \over {\bf e}(N_{Fix_{g,\mu_1,\mu_2,\dots,\mu_m}/\overline{\cal M}_{g,n}})}
\eeq
the localized integrals restricted to this subset of the fixed locus.

\begin{figure}
\includegraphics[width=15cm]{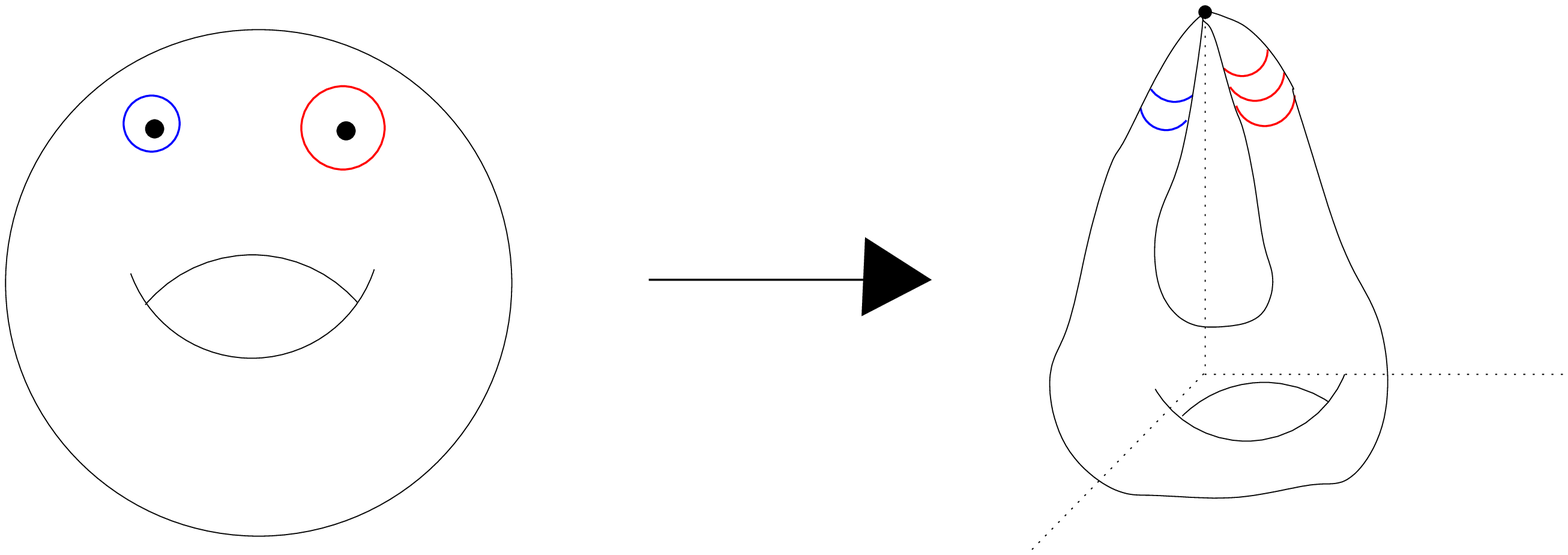}\\
  \caption{Example of stable map contributing to $A_{2,3}^{(1)}$. The blue (resp. red) cycle on the
  worldsheet winds twice (resp. three time) around the fixed point $p_1$ in the target space.}\label{figclosedtoricexample}
\end{figure}

Remark that these amplitudes are closely related to "open" amplitudes which are integrals over the
moduli space of maps from open worldsheets to $\CYX$ with fixed Brane boundary conditions by
removing a small circle from the closed surface around the marked points.

Let $\Sigma_g$ be a genus $g$ surface with $n$ marked points $z_1,\dots,z_n$ embedded into $\CYX$ thank to
a stable map $m_{g,n} \in Fix_{g,\mu}$ with $l(\mu)=n$ and $z_1,\dots,z_n$ all mapped to the fixed point
$p_1:=(X_1=\infty,X_2=0,X_3=0)$ of the compactification of $\CYX$.

For $r\in \mathbb{R}$ large enough, the pull-back of the circle $(|X_1|=r,X_2=0,X_3=0)$  by $m_{g,n}^{-1}$
has $n$ connected components winding $\mu_i$ times around $z_i$ respectively. Thus, removing
the pull-back of the discs $(|X_1|>r,X_2=0,X_3=0)$  by $m_{g,n}^{-1}$ from $\Sigma_g$ leaves us with an
open surface $\Sigma_{g,n}$ with $n$ boundaries whose embedding in $\CYX$ by $m_{g,n}$ is stable under the Torus action
with its boundaries mapped to the Brane $(|X_1|=r,X_2=0,X_3=0)$. This mapping between stable open surfaces and
marked closed ones is bijective up to symmetry factor and allows to get
$$
A_{\mu}^{(g)}\propto {\cal{N}}_{\mu}^{(g)}(r) := \sum_{Fix_{g,\mu}(r)} \int_{m_{g,\mu} \in Fix_{g,\mu}(r)} {i^* m_{g,\mu}\over {\bf e}(N_{Fix_{g,\mu_1,\mu_2,\dots,\mu_m}/\overline{\cal M}_{g,n}})}
$$
where $Fix_{g,\mu}(r)$ stands for the fixed locus of the moduli space of maps from an open surface with $l(\mu)$
boundaries mapped to the brane $(|X_1|=r,X_2=0,X_3=0)$. ${\cal{N}}_{\mu}^{(g)}(r)$ are exactly the volume
computed in the preceding sections.

\br
One can invert this procedure by recapping the open surface by a disc (or rather a semi-infinite cylinder). This gives the precise relation between the open and closed amplitudes. This is the relation \ref{eqrecap}.
\er

It is also interesting to remark that in this picture, the length of the boundaries of the open surfaces are
quantized once $r$ is fixed: they are labeled by the winding numbers.
The Laplace transform considered for the definition of the generating functions of these numbers in section \ref{secexpansions}
are thus replaced by discrete Laplace transforms with respect to the winding numbers\footnote{These transforms are
the one used in \cite{EMS,Chen,Zhou1,Zhou2} for the computation of Hurwitz numbers and Gromov-Witten invariants of $\mathbb{C}^3$.}.

\subsection{Cut-and-join, spectral curve and mirror map}

Following Kontsevich \cite{kontsevich}, one can describe the fixed locus of the moduli space of maps
$Fix_{g,\mu_1,\mu_2,\dots,\mu_k}$ rather
explicitly by associating a decorated graph to each component of this fixed locus. In particular, the case
$g=0$, $\mu_1 = \{d\}$ and $\mu_i = 0$ for $i\neq 1$ maps to the enumeration of rooted trees with labeled edges.
The integrals of the Euler classes on each component can then be performed explicitly (see \cite{Florea} for the framed
vertex or \cite{bookmirror} for the general Toric CY 3-fold case) and one ends up with a problem of enumeration
of trees with fixed weight.

This procedure gives the disc amplitudes. The generating function for these disc amplitudes was proved (at least physically \cite{Agvafa}) to be closely related to the
superpotential of the mirror theory: the problem of enumeration of trees can be solved by induction on the winding
number $d$; the generating function $W_1^{(0)}(x)$ for these numbers is thus a "tree function" solution to an
equation of the type $ H(\ee{x},\ee{W_1^{(0)}(x)})=0$ which is the equation of the singular locus of the B-model
target space.

Since the disc amplitudes are the only unknowns of the topological recursion, once we know the disc amplitudes, all the
other amplitudes can be computed by the topological recursion.
As explained in the preceding section, the generating function for the disc amplitudes defines the spectral curve. One thus recovers that the Gromov-Witten invariants of some A-model are given by the symplectic invariants of the corresponding B-model target space in the coordinates associated to A-Branes under study.

\bigskip

Many questions arise from this approach and have to be addressed in order to clarify the combinatorics
underlying the topological recursion. An important
one is to make clear how the choice of a torus action maps to the choice of parameterization of the
B-model target space (or of the spectral curve in our language).

Given a Torus action used for the localization method, there is no ambiguity in the determination of the integrals
over the moduli space of maps: it fixes both the polarization and the framing by fixing the Torus action. In the
simplest case of a $\mathbb{C}^3$ target space, the map between the choice of a Torus action and
a parametrization of the spectral curve has already been performed (see for example \cite{Florea} or \cite{Zhou1}) and the framings
of the corresponding Toric vertex.

For a general toric CY target space, the theory of the topological vertex \cite{mathematicalvertex}
 allows us to follow the same procedure and associate a parameterized curve to a given torus
action, i.e. it maps a foliation of the worldsheet to a parameterization of the spectral curve.
The topological recursion follows from a particular local parameterization of the spectral curve.
It would be interesting to see if this local parametrization of the spectral curve can be mapped to a local
$S^1$ action on the target space, and thus, to a local foliation of the worldsheets.

\section{Conclusion}

First, let us say again that all the construction presented in this paper is semi-heuristic and many details need to be further studied and made precise in order to become really rigorous.
It can't be seen as a proof of the "remodeling the B-model" formalism, but rather as an intuitive geometric understanding of it.
One could expect to find a rigorous proof of BKMP by using generalized cut and join equations, as we suggest in section \ref{secexpansions}, this is more or less what was done for Hurwitz numbers in \cite{EMS} and for $\CYX=\mathbb C^3$ by Zhou \cite{Zhou1,Zhou2} and Chen \cite{Chen}.

\subsubsection*{Reverse engineering}

Our approach is a reverse engineering of the topological recursion.
In fact, knowing that a string theory satisfies the topological recursion, and knowing that the topological recursion depends only on the data of a spectral curve, we tried to reconstruct the string theory from the spectral curve.
The topological recursion clearly implies that every worldsheet should be decomposable in a unique way into propagators and cylinders. Therefore, there is no really other choice than the theory we presented here.

\medskip

The main drawback of that reverse engineering approach, is that, given a string theory, it is not so straightforward to recover the spectral curve associated to it.

In the context of topological B-strings in toric CY 3-fold target spaces, the spectral curve was found from mirror symmetry \cite{bookmirror}, and the "remodeling the B-model" idea of BKMP \cite{BKMP} is based on that.

\medskip
For other models, it would be interesting to see how our method applies in practice.
For instance, starting from the Lambert spectral curve, how do we see that our foliated flat worldsheets are realizing branched coverings of the sphere.

\smallskip
Also for matrix models, the spectral curve is well known, it is algebraic, it is related to the Toda chain integrable system, and it is known that it enumerates discrete surfaces (called "maps" by combinatorists).
It would be interesting to see how our flat coordinates foliate discrete surfaces in that context.
In other words, what are the horizontal and vertical trajectories on the discrete surfaces ?

\medskip
As pointed out in the preceding section, for the works on Hurwitz coverings and matrix models, the spectral curve
seems to always come from a cut-and-join procedure (or Tutte's equations for the matrix models) recursively defining
the disc amplitudes. The latter takes the form of an induction of a set of trees whose weights depend on the propagator
amplitude, i.e. the possible singularities of the chosen hamiltonian fibration. The study case by case of this general
statement would be very interesting.

\subsection*{Flat coordinates and flat surfaces}

In fact, this idea of foliating worldsheets with the help of the flat connection of an integrable system was already used in string theory. For instance in \cite{Dorey}, Dorey et al. used a flat connection to parametrize bare cylinders. They didn't consider higher topologies.

\bigskip

Let us also mention the link with the theory of flat surfaces developed by Zorich et al.
It is clear that our flat surfaces, can be realized as fundamental polygonal domains of a plane modulo the lattice, i.e. as a polygonal (embedded in the Jacobian, see fig.\ref{figmvtJac}), with some opposite sides glued together.
There is an important literature on that theory of flat surfaces, see \cite{Zorich}, and it would be interesting to clarify the link with our  approach.

\figureframex{12}{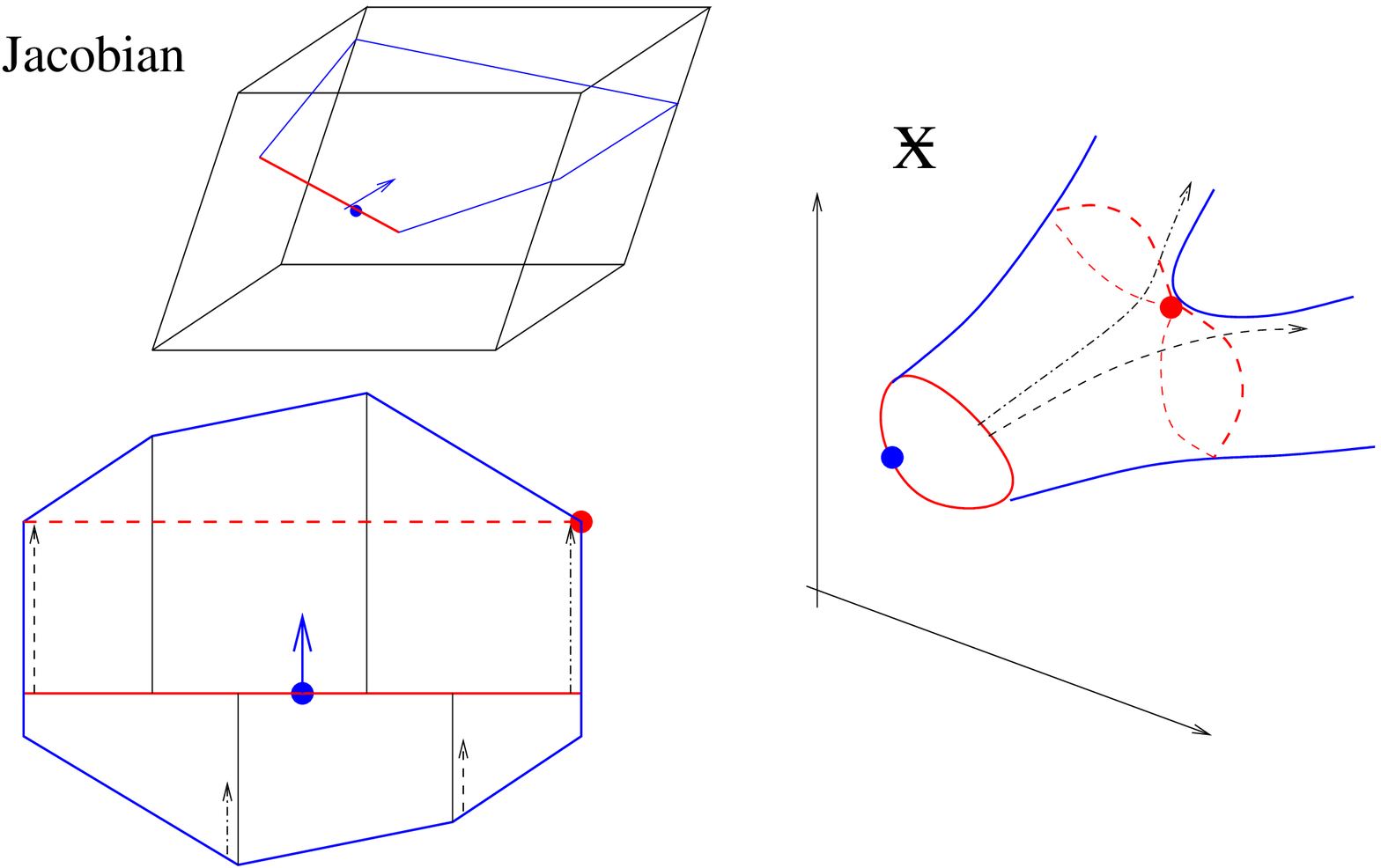}{\label{figmvtJac} The linear constant motion in the periodic Jacobian generates a flat surface obtained by gluing opposite sides of a polygon. The boundary at time $t=0$ has the topology of a circle, but at some critical time, it ceases to be a circle, there is a branching.
The image of this constant motion, produces a Riemann surface in the target space $\CYX$.}

\subsubsection*{Integrable systems}

The theory of Hitchin's systems, relies on a foliation of moduli spaces, with the orbits of an integrable system. The foliation concerns the moduli space itself, and not the worldsheets, however, it is clear that  there is a link between those two approaches ( a foliation of worldsheets induces a foliation of moduli spaces), and it would be worth developing it. From this perspective, it would be interesting to
compare our approach with the description of topological string theories by Gerasimov and Shatashvili in \cite{gershat}.

\medskip
Another link between enumerative geometry problems and integrable systems arises through the Frobenius manifolds structure.
There also, it seems interesting to understand how our approach fits in the framework of Frobenius manifolds. In
particular, the relation between flat and canonical coordinates in the work of Dubrovin and al. \cite{Dubrovin}
seems to coincide with the transform expressing the expansion in powers of local coordinates near the branch points, in terms of the expansions in terms of KP times related to the expansion near the poles of $ydx$, see section \ref{secexpansions}.
This points towards the link between a change of torus action in the topological
string setup and the structure of Frobenius manifolds.

\subsubsection*{Other prospects}

In \cite{DV2d}, Dijkgraaf and Vafa showed that Koddaira-Spencer theory also satisfies the topological recursion, and thus there is a quantum field theory equivalent to this formalism.
Another way of seeing a quantum field theory, is through integrability. Integrable systems' correlation functions are determinants, and can be written as free fermions integrals. In particular, it would be interesting to
generalize Kostov's work on CFT description of matrix models \cite{Kos} to
realize the correlation functions studied in this paper as correlation functions of some associated conformal field
theory on a Riemann surface with insertion of Twist operators associated to the branch points.

Finally, we considered only classical integrable systems in these notes. In \cite{CEM}, the topological recursion
has been generalized to quantum spectral curves $(\curve,x,y)$ with non-commuting  $x$ and $y$: $[y,x]=\hbar$.
It seems very likely that this quantization is related to Nekrasov's partition function for
equivariant theories \cite{Nekr1}. Generalizing
the method developed here in this non-commutative context could give a new geometrical point of view
on this partition function as well as cut-and-join like formulas for equivariant theories with higher
dimensional torus actions.
This could also lead to a link with the AGT conjecture \cite{AGT}.

\medskip

Also, let us mention that here we considered only "closed" strings, ending on D-branes, with Dirichlet boundary conditions, i.e. horizontal trajectories.
It seems easy to extend our method to worldsheets bounded also by vertical trajectories, which thus correspond to Von Neumann boundary conditions, and which realize open strings.
It seems that one could easily extract a topological recursion formula for open strings as well.

\section*{Aknowledgments}

We would like to thank O. Babelon, V. Bouchard, M. Mari\~no, S. Pasquetti for useful and fruitful discussions on this subject, and particularly M. Mulase for his advice and encouragements.
This work is partly supported by the
ANR project Grandes Matrices Al\'eatoires ANR-08-BLAN-0311-01, the European Science
Foundation through the Misgam program, and the Quebec government with the FQRNT.

\end{document}